# Google Scholar, Microsoft Academic, Scopus, Dimensions, Web of Science, and OpenCitations' COCI: a multidisciplinary comparison of coverage via citations

Alberto Martín-Martín [1] , Mike Thelwall [2] , Enrique Orduna-Malea [3] , Emilio Delgado López-Cózar[1]

## Abstract (also in Spanish, Chinese)

**Introduction**
New sources of citation data have recently become available, such as Microsoft Academic, Dimensions, and the OpenCitations Index of CrossRef open DOI-to-DOI citations (COCI). Although these have been compared to the Web of Science (WoS), Scopus, or Google Scholar, there is no systematic evidence of their differences across subject categories.

**Methods**
In response, this paper investigates 3,073,351 citations found by these six data sources to 2,515 English-language highly-cited documents published in 2006 from 252 subject categories, expanding and updating the largest previous study.

**Results**
Google Scholar found 88% of all citations, many of which were not found by the other sources, and nearly all citations found by the remaining sources (89%-94%). A similar pattern held within most subject categories. Microsoft Academic is the second largest overall (60% of all citations), including 82% of Scopus citations and 86% of Web of Science citations. In most categories, Microsoft Academic found more citations than Scopus and WoS (182 and 223 subject categories, respectively), but had coverage gaps in some areas, such as Physics and some Humanities categories. After Scopus, Dimensions is fourth largest (54% of all citations), including 84% of Scopus citations and 88% of WoS citations. It found more citations than Scopus in 36 categories, more than WoS in 185, and displays some coverage gaps, especially in the Humanities. Following WoS, COCI is the smallest, with 28% of all citations.

**Conclusions**
Google Scholar is still the most comprehensive source. In many subject categories Microsoft Academic and Dimensions are good alternatives to Scopus and WoS in terms of coverage.

---

[1] Facultad de Comunicación y Documentación, Universidad de Granada, Granada, Spain.
[2] Statistical Cybermetrics Research Group, School of Mathematics and Computer Science, University of Wolverhampton, Wolverhampton, UK.
[3] Universitat Politècnica de València, Valencia, Spain.

✉ Alberto Martín-Martín
albertomartin@ugr.es

# 1. Introduction

## 1.1 Timeline

The first scientific citation indexes were developed by the Institute for Scientific Information (ISI). The Science Citation Index (SCI) was introduced in 1964, and was later joined by the Social Sciences Citation Index (1973) and the Arts & Humanities Citation Index (1978). In 1997, these citation indexes were moved online under the name "Web of Science". Recently, these citation indexes, along with some new ones such as the Conference Proceedings Citation Index, the Book Citation Index, and the Emerging Sources Citation Index, were rebranded as the "Web of Science Core Collection" (from now on, WoS). The availability of this data was essential to the development of quantitative studies of science as a field of study (Birkle et al., 2020).

In November 2004, two new academic bibliographic data sources that contained citation data were launched. Like WoS, Elsevier's Scopus is a subscription-based database with a selective approach to document indexing (documents from a pre-selected list of publications). A few weeks after Scopus, the search engine Google Scholar was launched. Unlike WoS and Scopus, Google Scholar follows an inclusive and automated approach, indexing any seemingly academic document that its crawlers can find and access on the web, including those behind paywalls through agreements with their publishers (Van Noorden, 2014). Additionally, Google Scholar is free to access, allowing users to access a comprehensive and multidisciplinary citation index without charge.

In 2006, Microsoft launched Microsoft Academic Search but retired it in 2012 [4] (Orduña-Malea et al., 2014). In 2016, Microsoft launched a new platform called Microsoft Academic, based on Bing's web crawling infrastructure. Like Google Scholar, Microsoft Academic is a free academic search engine, but unlike Google Scholar, Microsoft Academic facilitates bulk access to its data via an Applications Programming Interface (API) (Wang et al., 2020).

In 2018, Digital Science launched the Dimensions database (Hook et al., 2018). Dimensions uses a freemium model in which the basic search and browsing functionalities are free, but advanced functionalities, such as API access, require payment. This fee can be waived for non-commercial research projects.

Also in 2018, the organization OpenCitations, dedicated to developing an open research infrastructure, released the first version of its COCI dataset (OpenCitations Index of CrossRef open DOI-to-DOI citations). The citation data in COCI comes from the lists of references openly available in CrossRef (Heibi et al., 2019). Until 2017, most publishers did not make these references public, but the Initiative for Open Citations (I4OC), launched in April 2017, has since convinced many publishers to do so. The rationale is that citation data should be considered a part of the commons and should not be only on the hands of commercial actors (Shotton, 2013, 2018). At the time of writing, 59% of the 47.6 million articles with references deposited with CrossRef have their references open [5]. However, some large publishers, such as Elsevier, the American Chemical Society, and IEEE have not yet agreed to opening their lists of references. Thus, COCI's only partially reflects the citation relationships of documents recorded in CrossRef, which now covers over 106 million records (Hendricks et al., 2020).

---

[4] https://web.archive.org/web/20170105184616/https:/academic.microsoft.com/FAQ
[5] https://i4oc.org/

The new bibliographic data sources are changing the landscape of literature search and bibliometric analyses. The openly available data in Microsoft Academic Graph (MAG) has been integrated into other platforms, significantly increasing their coverage (Semantic Scholar, Lens.org). There are still some reuse limitations, such as that the current license of MAG (ODC-BY) requires attribution, which apparently precludes it from being able to be integrated into COCI (which uses a CC0 public domain license). This openness is nevertheless an advance on the previous situation, in which most citation data was either not freely accessible (WoS, Scopus), or free but with significant access restrictions (Google Scholar). At this point, citation data is starting to become ubiquitous, and even owners of closed bibliographic sources, such as Scopus, are beginning to offer researchers options to access their data for free [6].

Other citation indexes have been developed within various academic platforms, but these are not analysed in this study, for various reasons:

- CiteSeerX [7], from Penn State University, indexes documents in the public web and not those that are only found behind paywalls (Wu et al., 2019).
- ResearchGate [8] generates its own citation index based on the full text documents that its crawler finds on the Web and those that its users upload to the platform. However, the platform does not offer a way to extract data in bulk, and it is difficult to use web scraping to obtain data because the complete list of citations to an article cannot be easily displayed.
- Lens.org [9] integrates coverage from Microsoft Academic, CrossRef, PubMed, and a number of Patent datasets. It is not included in this analysis because two of its main sources (Microsoft Academic and CrossRef) are already included.
- Semantic Scholar [10] originally focused on Computer Science and Engineering. Later it expanded to include Biomedicine, and recently it has integrated multidisciplinary coverage from Microsoft Academic (which is also the reason why we decided not to analyse it).
- There are also several regional or subject-specific citation indexes, which only index documents published by journals and/or researchers who work in a specific country or region, or in specific topics. Given their specific scope these are not easily comparable to sources with a worldwide and/or multidisciplinary coverage.

### 1.2. Previous analyses of coverage

Document coverage varies across data sources (Ortega, 2014), and studies that analyse differences in coverage can inform prospective users about the comprehensiveness of each database in different subject areas. For citation indexes, greater coverage should equate to higher citation counts for documents, if citations can be extracted from all documents. Coverage is not the only relevant aspect that should be considered when deciding which data source should be used for a specific information need (e.g., literature search, data for bibliometric analyses). Other aspects such as functionalities to search, analyse, and export data, as well as transparency and cost, are also relevant, but not analysed here. Some of these aspects are analysed by Gusenbauer & Haddaway (2020).

---

[6] https://www.elsevier.com/icsr/icsrlab
[7] https://citeseerx.ist.psu.edu/index
[8] https://www.researchgate.net/
[9] https://www.lens.org/
[10] https://www.semanticscholar.org/

### 1.2.1 The veterans: WoS, Scopus, and Google Scholar

As the longest-running platforms, many studies have analysed the differences in coverage and citation data between WoS, Scopus, and Google Scholar. WoS covers over 75 million records in its Core Collection (which includes its main citation indexes), and up to 155 million records when other regional and subject-specific citation indexes are included (Birkle et al., 2020). Scopus claims to cover over 76 million records (Baas et al., 2020). Google Scholar does not disclose official figures about its coverage (Van Noorden, 2014), but the most recent independent studies have estimated that it covers well over 300 million records (Delgado López-Cózar et al., 2019; Gusenbauer, 2018). At this point most studies agree that Google Scholar has a more comprehensive coverage than Scopus and WoS, and includes the great majority of the documents that they cover. However, the relatively low quality of the metadata available in Google Scholar and the difficulty to extract it make it challenging to use Google Scholar data in bibliometric analyses (Delgado López-Cózar et al., 2019; Halevi et al., 2017; Harzing, 2016; Harzing & Alakangas, 2016; Martín-Martín et al., 2018; Moed et al., 2016).

### 1.2.2 Microsoft Academic

Microsoft Academic has been recently reported to cover over 225 million publications (Wang et al., 2020). Harzing carried out an analysis of her own publication record and the publication records of 145 academics in five broad disciplinary areas (Harzing, 2016; Harzing & Alakangas, 2017a, 2017b). Microsoft Academic found more of her own publications than Scopus or WoS. For the sample of publications by 145 academics, Microsoft Academic provided higher citation counts than both Scopus or WoS in Engineering, Social Sciences, and the Humanities, and similar figures in Life Sciences and Sciences. Google Scholar reported the highest citation counts in all disciplines.

Hug & Brändle (2017) also analysed the coverage of Microsoft Academic and compared it to Scopus and WoS. Based on publications included in the repository of the University of Zurich as a case study, Microsoft Academic had wider coverage of non-article documents than Scopus and WoS, while Scopus had a slightly lower coverage of journal articles than Microsoft Academic. Microsoft Academic showed similar biases to Scopus and WoS against non-English publications and publications in the Humanities. Haunschild et al. (2018) analysed a subset of the same sample used in the previous study (25,539 papers also covered by WoS) and found that 11% had no associated cited references in Microsoft Academic, while in WoS the same papers had associated cited references. However, for publications with less than 50 associated references in WoS (24,788) the concordance correlation coefficient applied to the number of references found by each source was 0.68, indicating a strong tendency for them both to report the same number.

Thelwall (2017) analysed the citation counts of 172,752 articles in 29 large journals from various disciplines, and compared them to Scopus citation counts and Mendeley reader counts. For articles published between 2007 and 2017, Microsoft Academic found slightly more citations than Scopus overall, and significantly more than Scopus for documents published in 2017. In subsequent studies, Thelwall (2018a) found that Microsoft Academic did find earlier citations to recently published articles when compared to Scopus. Kousha & Thelwall (2018) studied the coverage and citation counts of books in Microsoft Academic and Google Books by analysing a sample of book records extracted from the Book Citation Index (BKCI) in WoS. They found 60% of the books in their sample overall, but this percentage was lower in some categories of the Humanities and Social Sciences. Citation counts in Microsoft Academic were higher than in BKCI in 9 out of 17 fields during 2013-2016. Kousha et al. (2018) analysed whether Microsoft Academic was able to find early citations of in-press articles using a sample of 65,000 in-press articles from 2016-2017, and found that Microsoft Academic was able to find 2-5 times as many citations as Scopus. This was mostly because Microsoft Academic (like Google Scholar) merges preprints (and the citations these receive) with their subsequent in-press versions, and because Microsoft Academic covers repositories such as arXiv.

Visser et al. (2020) carried out a large-scale comparison of WoS, Scopus, Dimensions, Microsoft Academic, and CrossRef by matching the entire collection of documents in each source. They found that Microsoft Academic was the source with the largest coverage overall, and the one with the higher overlap with Scopus documents (81% of Scopus documents were found in Microsoft Academic). Some of the documents in Microsoft Academic were not of a scientific nature. Microsoft Academic was not able to detect 12.7% of the citations found by Scopus after adjusting for coverage differences.

### 1.2.3 Dimensions

Dimensions covers over 105 million publications, as well as other kinds of records such as grants data, clinical trials, patents, and policy documents (Herzog et al., 2020).

Orduña-Malea & Delgado-López-Cózar (2018) analysed several small samples of journals, documents and authors in the field of Library & Information Science using Dimensions, and compared the data to Scopus and Google Scholar. Dimensions provided slightly lower citation counts than Scopus. Thelwall (2018c) analysed a random sample of 10,000 Scopus articles from 2012, finding that Dimensions covered the great majority of articles with a DOI (97%) and high correlations between citation counts in the two sources (median of 0.96 across narrow subject categories).

Harzing (2019) analysed coverage of Dimensions and CrossRef, and compared it to the coverage in WoS, Scopus, Google Scholar, and Microsoft Academic using her own publication and citation record, as well as that of six top journals in Business & Economics. CrossRef and Dimensions had similar or better coverage of publications, and similar citation counts to those in WoS and Scopus, but still substantively lower than Google Scholar and Microsoft Academic.

Visser et al. (2020) found that Dimensions had a substantially higher coverage than Scopus and WoS, which heavily relied on data from CrossRef. After computing the overlap in coverage between Dimensions and Scopus, they found that overall, Dimensions covered 78% of the documents available in Scopus (35.1 million out of 44.9 million documents in Scopus). They also analysed the accuracy and completeness of citation links, finding that, after adjusting for coverage differences, there were 489.7 million citations found by both sources (percentage of full overlap: 83%), 73.2 million only found by Scopus, and 25.8 million only found by Dimensions.

### 1.2.4 COCI

COCI has detected over 624 million citation relationships involving over 53 million documents (Peroni & Shotton, 2020). The citations recorded in this source are only a fraction of the citations that have actually occurred among the documents covered by CrossRef, because some publishers that deposit lists of references or CrossRef have not agreed to make them available, and other publishers and preprint servers do not deposit any references in CrossRef or do it only for some document types (Shotton, 2018; van Eck et al., 2018). Huang et al. (2020) used citation data from COCI and bibliographic data from WoS, Scopus and Microsoft Academic to test the robustness of university rankings created with these different sources, and concluded that despite its lack of comprehensiveness COCI is already a viable data source for cross comparisons at the system level.

## 1.3 Objective

The citation index coverage studies published so far have analysed a heterogeneous variety of samples of documents, disciplines, and data sources. In response, this paper reports a systematic comparison of coverage of six data sources (Google Scholar, Microsoft Academic, Scopus, Dimensions, WoS, and COCI [11]) across 252 subject categories using a relatively large sample of citations. This allows comparisons across a large number of disciplines for the most widely used bibliographic data sources. This study expands and updates a previous analysis of Google Scholar, Scopus and WoS (Martín-Martín et al., 2018). The main research question that drives this is investigation is:

> RQ. How much overlap is there between Google Scholar, Microsoft Academic, Scopus, Dimensions, WoS, and COCI in the citations that they find to academic documents and does this vary by subject?

# 2. Methods

## 2.1. Direct coverage comparison vs. comparison of citations

The most direct method to compare document coverage across different data sources would be to obtain a complete list of all documents covered by each source, match the documents across databases, and report the size of the overlaps (Visser et al., 2020). This is not possible here because of access restrictions. For example, Scopus and WoS charge for this kind of access and Google Scholar does not share its database.

Because of these restrictions, studies analysing coverage differences across bibliographic data sources often use an alternative method: they select a seed sample of documents that are known to be covered by all the data sources under analysis, and then they compare the list of citing documents that each data source is able to find for each of the seed documents (Martín-Martín et al., 2018). The rationale of this method is that if data source *A* is not able to find a citation that data source *B* has found, the reason must be that the citing document is not covered by data source *A*. This assumes that all data sources are equally effective in detecting citation relationships. In fact, each data source has its own (usually secret) citation detection algorithms, and small discrepancies in citation data across databases exist even when removing the factor of differences in coverage (van Eck & Waltman, 2019; Visser et al., 2020). Furthermore, it is known that bibliographic databases do not always have access to cited reference lists for all the documents they cover, which also affects the citations they can detect. For example, reference lists are only available in a fraction of the documents indexed in CrossRef, so an analysis of the citations detected in this source does not accurately reflect the true size of the bibliographic database. Other sources, especially academic search engines, are also affected by this issue to some degree [12]. Lastly, academic documents that do not cite and are not cited by other documents cannot be detected by this type of analysis. Therefore, results from studies that analyse citations to identify relative differences in the sizes of bibliographic databases are likely to be affected by these confounding factors.

---

[11] In the case of COCI, the results cannot reflect the full coverage of CrossRef given the incomplete availability of reference lists in this source. Nevertheless, including it in the analysis will inform us of what proportion of citations are currently available in the public domain.

Of the six data sources that are analysed in this study, only two (Microsoft Academic and COCI) offered free and unrestricted access to the complete list of documents (or citation relationships in the case of COCI) that they covered at the time of data collection, although Dimensions now also offers this to researchers. To include all data sources in this study in a comparable way, the alternative method (selection of seed sample and analysis of citations) was used to discover relative coverage differences among data sources across subject categories. Since citation extraction discrepancies seem likely to be small compared to coverage differences, the results should also be useful to detect differences in coverage between sources.

## 2.2. Selection of seed sample

The sample of citations analysed in this paper was taken from a seed sample of highly-cited documents: those listed in Google Scholar's *Classic Papers* product [13] (GSCP). This sample comprises the top 10 most cited documents published in 2006 according to Google Scholar in each of 252 subject categories (except *French Studies*, which has only 5 documents). The 252 subject categories are also assigned to one or more of 8 broad subject areas. The seed sample contains a total of 2,515 highly-cited documents. For more information on GSCP, see Orduna-Malea et al. (2018).

This seed sample was considered useful for the purpose of this study, as it is the only sample of documents in Google Scholar for which an article-level subject classification is available. At the time of data collection, no other sample of documents with an article-level classification was readily available to us, and a sample with these characteristics was considered superior to the journal-level classification schemes that are used in sources such as Web of Science and Scopus. Additionally, being aware of the difficulties that extracting data from Google Scholar entail (Else, 2018), the election of a sample of documents that were known to be highly cited also guaranteed a high efficiency in the citation extraction process (each request to Google Scholar retrieved the maximum amount of records that the search engine displays per page).

This study analyses the complete list of documents that cite this seed sample, as reported in a variety of citation indexes (Google Scholar, Microsoft Academic, Scopus, Dimensions, Web of Science, and COCI). In this study, they are called citing documents, or more simply, citations. Thus, this study follows the same approach as Martín-Martín et al. (2018).

## 2.3. Collection of citation data

Each of the 2,515 highly-cited documents were searched on Google Scholar, Microsoft Academic, Scopus, Dimensions, WoS, and COCI (Table 1). For each seed document found in a data source, the list of citing documents was extracted, as described below.

---

[12] Visser et al. (2020) found that a large number of citations missing from Microsoft Academic were caused by missing reference lists in the citing documents. As far as we know no study has analysed how many missing citations in Google Scholar are caused by missing reference lists.
[13] https://scholar.google.com/citations?view_op=list_classic_articles&hl=en&by=2006

The searches and data extraction were carried out in May and June 2019 (i.e., not reusing the data from the previous paper).

Google Scholar has no data exporting capabilities in its web interface and no API. Instead, a custom web scraper was used to extract the list of citing documents for each highly-cited document in the seed sample (Martín-Martín, 2018). CAPTCHAs were solved manually when they appeared.

Google Scholar provides up to 1,000 results per query. In order to download the complete list of citing documents for those with more than 1,000 citations, queries were split by the publication year of the citing documents. Using this method, we were able to download most of the citing documents available in Google Scholar: for 2,429 (96.5%) seed documents, we were able to extract a list of citing documents, amounting to at least 98% of the total citation counts reported by Google Scholar for these seed documents. In eight cases (extremely highly-cited seed documents), splitting queries by publication year was not enough to find all possible citing documents, and in these cases the number of citing documents extracted from Google Scholar was lower than 75% of the reported Google Scholar citation counts. This disadvantages Google Scholar in comparison to the other sources, for which all citing documents could be extracted. 2,689,809 citations were extracted from Google Scholar.

The metadata provided by Google Scholar is limited (Delgado López-Cózar et al., 2019). For example, Google Scholar does not provide the DOI of a document, which is very useful for document matching across data sources, and therefore relevant to our study. To enrich the limited metadata provided by Google Scholar, we followed several approaches. First, given that most of the citing documents from Google Scholar had already been analysed (Martín-Martín et al., 2018), we matched the newly extracted list of citing documents to the data from the previous study, and retrieved all the enriched metadata that was available in the dataset used for the 2018 study. Next, for all the citing documents that could not be matched in the previous step (mostly newer citations), metadata was extracted from the HTML Meta tags in the landing page of each citing document, and with public metadata APIs when a CrossRef or DataCite DOI could be found. These methods produced a DOI for 62.9% of all Google Scholar citations.

To collect citation data from Microsoft Academic, the Academic Search API[14] was used. This API is free with a limit of 10,000 transactions per month. At the moment of data collection, this API did not facilitate searching directly by DOI (Thelwall, 2018b). For this reason, every highly-cited seed document was first searched for by title. Once the seed document was retrieved and confirmed to be correct, new queries were submitted to retrieve the list of citing documents. Up to 1,000 citing documents per query could be extracted (seed documents with over 1,000 citations required more than one query to extract all citations). For each citing document, the Microsoft Academic internal Id, as well as the DOI, the document title, the list of authors, the publication year, the language, and the citation counts, were retrieved. 1,840,702 citations were extracted from Microsoft Academic.

To collect citation data from Scopus, the exporting capabilities of the web interface were used. Each seed highly-cited document was searched in Scopus by DOI and title, and, if found, the list of citing documents was exported in csv format. Scopus allows 2,000 records per query to be exported. When seed documents had over 2,000 citations, the alternative email service was used, which allows 20,000 records to be exported. No document in the seed sample had more than 20,000 citations in Scopus. 1,738,573 citations were extracted from Scopus.

---

[14] https://msr-apis.portal.azure-api.net/docs/services/academic-search-api

To collect citation data from Dimensions, its API was used, which is free for research [15]. The Dimensions API allows searching by DOI. Therefore, all seed highly-cited documents were searched for using their DOI, and, when unavailable, by their title. Once all the seed documents had been identified in Dimensions, the API was also used to extract the list of citing documents. For each citation, the basic bibliographic information (DOI, title, authors, publication year, source, document type) was recorded. 1,649,162 citations were extracted from Dimensions.

To collect citation data from WoS, the web interface was used. All citation indexes in WoS Core Collection were included in the analysis, including the Emerging Sources Citation Index (from publication year 2005 to the present). Each seed highly-cited document was searched by its DOI, and, when unavailable, by its title. The list of citing documents was then exported in batches of up to 500. The exported files were consolidated into a single table using a set of R functions (Martín-Martín & Delgado López-Cózar, 2016). 1,503,657 citations were extracted from WoS.

To collect citation data from COCI, the public API was used. The DOI of each seed highly-cited document was searched in order to retrieve the complete list of citing DOIs. 852,413 citation relationships were extracted from COCI.

*Table 1. Nº of seed highly-cited documents and citations found in each data source*

| Source | Seed documents* | | Citations |
|---|---|---|---|
| | N | % | |
| Google Scholar | 2,515 | 100 | 2,689,809 |
| Microsoft Academic | 2,500 | 99.4 | 1,840,702 |
| Scopus | 2,447 | 97.3 | 1,738,573 |
| Dimensions | 2,478 | 98.5 | 1,649,162 |
| WoS | 2,342 | 93.1 | 1,503,657 |
| COCI | 2,471 | 98.3 | 852,413 |

\* Due to the sample selection process, the figures related to the seed documents found in each data source cannot be used as evidence that Google Scholar has higher coverage than the other sources.

## 2.4. Analysis of citation data

To calculate citation overlaps across data sources, the citing documents from different data sources were matched. The matching process started with two data sources (WoS and Scopus), and the result was a full join of the two sources: a table containing all citations found both by WoS and Scopus, as well as the citations found only by one of the data sources. The resulting dataset was matched to the data obtained from another data source (Dimensions), and this process was repeated until all data sources were merged into a master list of citations (Table 2). The matching criteria are below, and are the same as previously used (Martín-Martín et al., 2018):

1. For each pair of data sources *A* and *B* and a seed highly-cited document *X*, all citing documents with a DOI that cite *X* according to *A* where matched to all citing documents with a DOI that cite *X* according to *B*.
2. For each of the unmatched documents citing *X* in *A* and *B*, a further comparison was carried out (except in the matching round where COCI data was integrated into the master table). The title of each unmatched document citing *X* in *A* was compared to the titles of all the unmatched documents citing *X* in *B*, using the restricted Damerau-Levenshtein distance (optimal string alignment) (Damerau, 1964; Levenshtein, 1966). The pair of citing documents which returned the highest title similarity (1 is perfect similarity) was selected as a potential match. This match was considered successful if either of the following conservative heuristics was met:
- o The title similarity was at least 0.8, and the title of the citing document was at least 30 characters long (to avoid matches between short, undescriptive titles such as "Introduction").
- o The title similarity was at least 0.7, and the first author of the citing document was the same in *A* and *B*.

---

[15] https://www.dimensions.ai/scientometric-research/

*Table 2. Rounds of the matching process*

| Matching round | Data sources being matched | Resulting dataset | Merged citations |
|---|---|---|---|
| 1st | WoS ⋈ Scopus | master_1 | 1,852,681 |
| 2nd | master_1 ⋈ Dimensions | master_2 | 1,990,862 |
| 3rd | master_2 ⋈ Microsoft Academic | master_3 | 2,263,896 |
| 4th | master_3 ⋈ COCI | master_4 | 2,273,067 |
| 5th | master_4 ⋈ Google Scholar | master_5 | 3,073,351 |

The matching criteria described above are intentionally conservative, so a match is only accepted when the two documents have very similar metadata. The analysis does not attempt to remove duplicate citations within the same data source, although Google Scholar and Scopus (and perhaps others) are afflicted by this issue (Orduna-Malea et al., 2017; van Eck & Waltman, 2019). In this study, if there are duplicate citations within the same data source only one of the instances will be linked to the same citation in other sources, while the rest will (erroneously) appear as unique citations. Therefore, the percentage overlaps between sources calculated are conservative estimates (i.e., they might be higher than reported here). A replication of the overlap analysis carried in Martín-Martín et al. (2018) for one subject category (Operations Management) showed that overlap figures are affected little when duplicates are identified and removed, however (Chapman & Ellinger, 2019).

Given that the objective is to detect relative differences in coverage across databases, to make comparisons as fair as possible the subset of citations that are considered in each comparison is adapted to include only citation relationships where the cited seed document is covered by all sources present in the comparison. For example, in a comparison of coverage across the six data sources analysed in this study (Table 1, top), only citations to the 2,319 seed highly-cited documents covered by all six data sources are considered. However, in pairwise comparisons, such as the Venn diagram that represents overlapping and unique citations in Google Scholar and Microsoft Academic (Figure 2A), the citations to the 2,500 seed highly-cited documents that are known to be covered by these two sources were analysed.

Data processing was carried out with the R programming language (R Core Team, 2014) using several R packages and custom functions (Dowle et al., 2018; Krassowski, 2020; Larsson et al., 2018; Martín-Martín & Delgado López-Cózar, 2016; van der Loo et al., 2018; Walker & Braglia, 2018; Wickham, 2016; Wilke, 2019). The resulting data files are openly available [16].

---
[16] https://osf.io/gnb72/ (2019 folder)

# 3. Results

## 3.1. Overall results (all subject categories)

### 3.1.1. Relative overlap

Overall, Google Scholar has the highest coverage, as it found 88% of all possible citations (2,918,105) to the 2,319 highly-cited documents in our sample that were covered by the six sources under analysis (Figure 1, first row). Microsoft Academic, Scopus, Dimensions and WoS found substantially fewer (60%-52% of all citations). COCI found only 28% of all possible citations.

In terms of relative overlaps between two data sources, larger data sources are able to find a vast majority of the citations found by the smaller data sources (Figure 1, row 2 through 6). Thus, Google Scholar found 89% of the citations in the second data source with the largest coverage (Microsoft Academic), and up to 94% of the citations in the smaller sources (WoS, COCI). On the other side of the spectrum, COCI, the smallest source, found between 30% and 51% of the citations found by the other sources (Google Scholar and Dimensions, respectively).

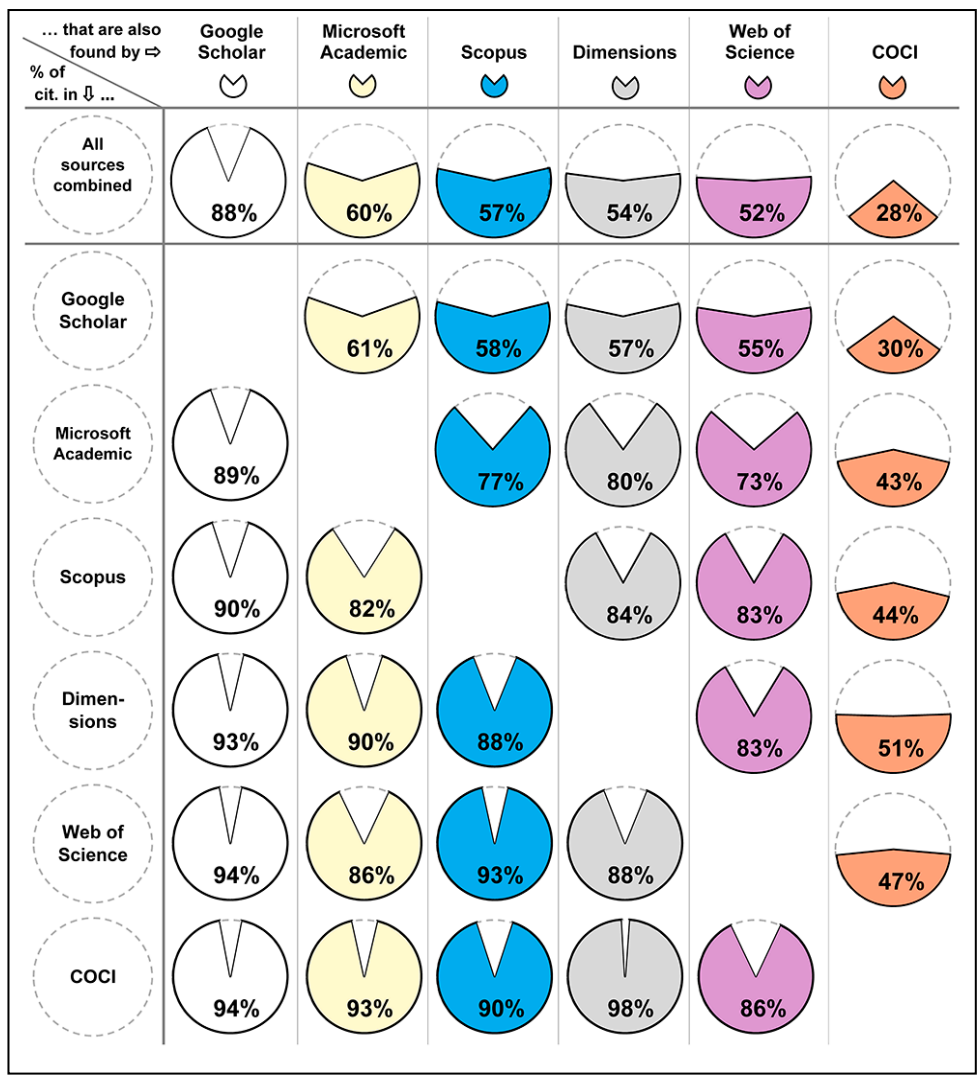

*Figure 1. Percentage of citations found by each database, relative to all citations (first row), and relative to citations found by the other databases (subsequent rows)*

For Microsoft Academic, Scopus, Dimensions, and WoS, the relative overlap between any two of these sources ranges from high (WoS found 73% of the citations found by Microsoft Academic) to almost full overlap (Dimensions found 98% of the citations available in COCI). Figure 1 shows that it is not always the case that the larger the source, the higher the proportion of citations from another source that it will be able to find. For example, Dimensions found 80% of the citations available in Microsoft Academic, while Scopus (larger than Dimensions) found 77%. The cause of this might be that both Microsoft Academic and Dimensions cover non-journal content, such as preprints, while Scopus does not. Scopus found 93% of the citations found by WoS, while Microsoft Academic (larger than Scopus) found 86%. Dimensions was able to find the highest proportion of COCI citations (98%) out of all the other sources (including Google Scholar).

### 3.1.2. Overlaps within the full set of citations

A quarter (26%) of all citations were found only by Google Scholar (Figure 2), 21% of the citations were found by the six sources, while 18% were found by all sources except COCI. The remaining 35% were found by combinations of four or less data sources, and the highest values were found in sectors that include Google Scholar and/or Microsoft Academic.

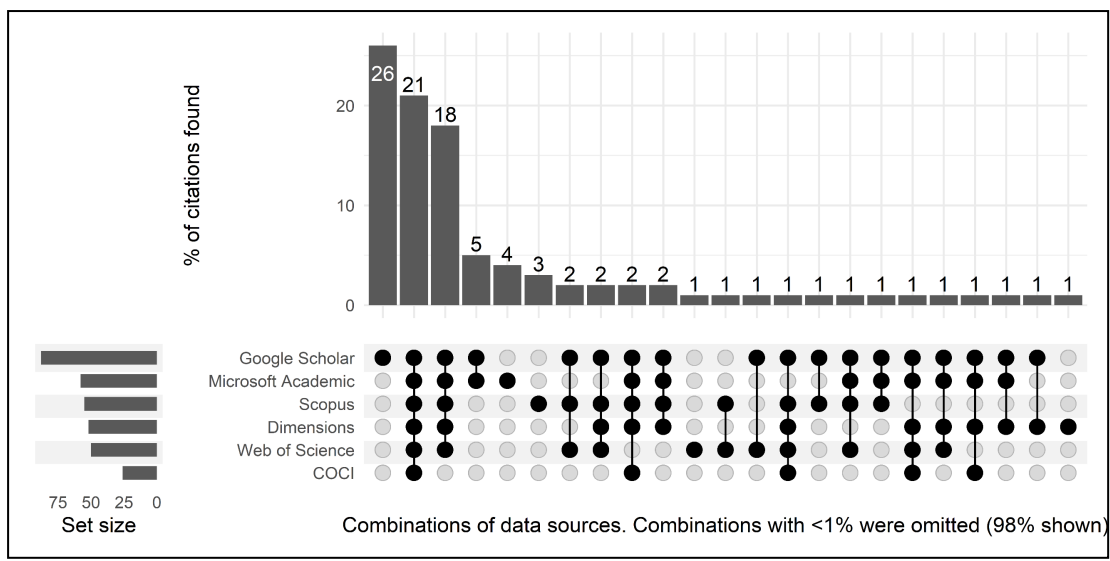

*Figure 2. Overlaps of citations found by Google Scholar, Microsoft Academic, Scopus, Dimensions, Web of Science, and COCI. Values expressed as percentages relative to N = 2,918,105 citations to 2,319 documents. Combinations with values below 1% are not displayed.*

When considering all possible pairwise combinations (Figure 3), the pairs of data sources that are most similar in terms of full citation overlap are Scopus/WoS (78% of all citations found by either were found by both), followed by Scopus/Dimensions (75%), Dimensions/WoS (75%), and Microsoft Academic/Dimensions (74%). Pairs that include Google Scholar or COCI display lower percentages of overlap: in the case of Google Scholar this is caused by the extra coverage in Google Scholar that is not found in the other sources, while for COCI the reason is the opposite.

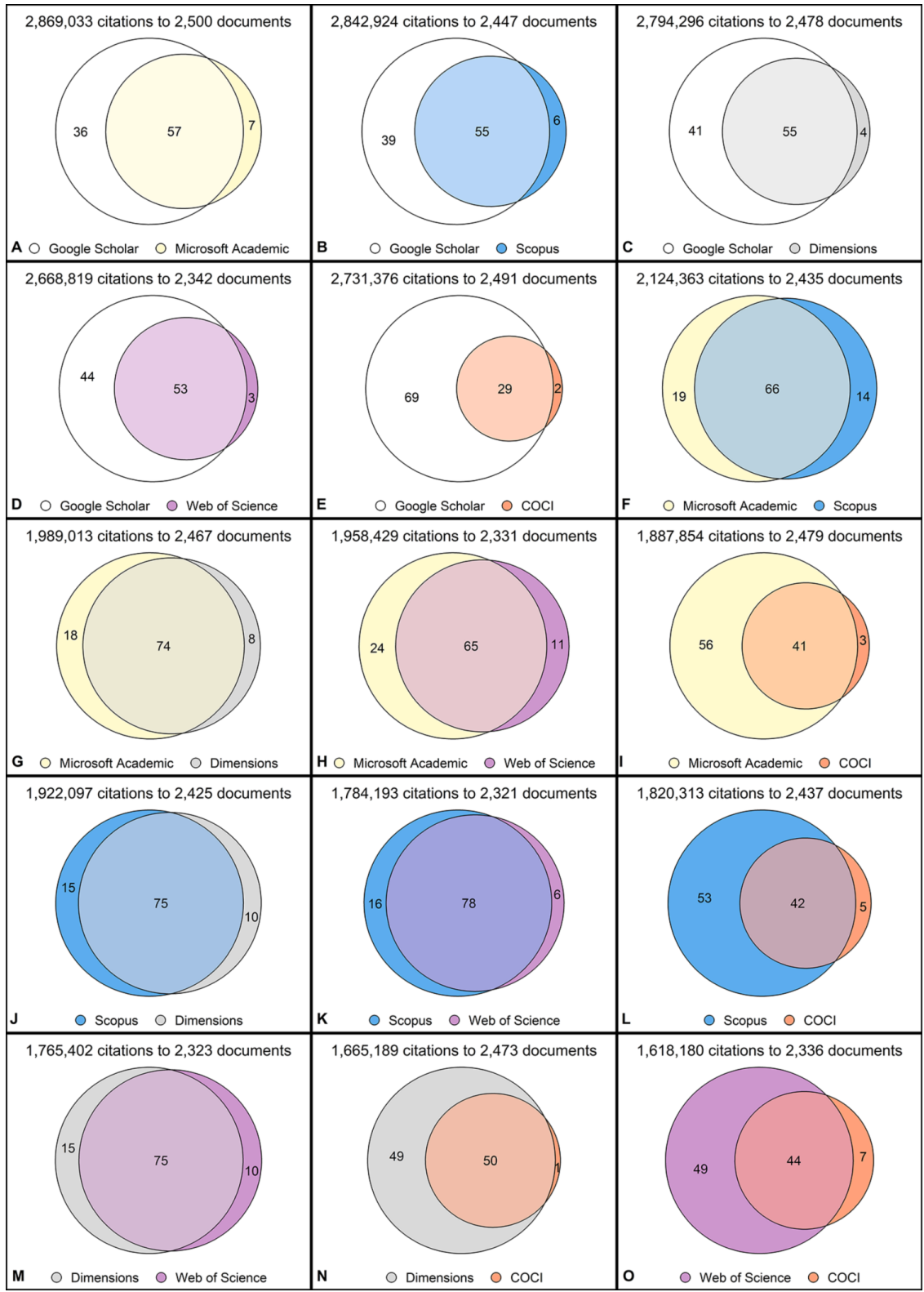

*Figure 3. Comparison of citing document overlaps between Google Scholar, Microsoft Academic, Scopus, Dimensions, Web of Science, and COCI (pairwise). Figures within Venn diagrams expressed as percentages.*

## 3.2. Analysis by subject areas and categories

### 3.2.1. Relative overlap

Disaggregating the data by broad subject areas provides a more detailed picture of the coverage of these sources. Google Scholar found the great majority of citations (85%-90%) in all eight subject areas (Table 3) and COCI found the fewest. COCI has differences in coverage across areas: in the Humanities and Social Sciences it found 18%-20% of all citations, while in the STEM areas (Science, Technology, Engineering, and Mathematics) it found a higher proportion of citations (27%-32%).

Between these two extremes, the other four sources (Microsoft Academic, Scopus, Dimensions, and WoS) tend to have similar coverage of each field, but differences between fields (Table 3). They have more comprehensive coverage for *Chemical & Material Sciences* (69%-72%), followed by *Life Sciences & Earth Sciences* (60%-68%). Conversely, their coverage is much lower in *Humanities, Literature & Arts* (25%-39%), *Social Sciences* (33%-47%) and *Business, Economics & Management* (29%-47%). Among these four, Microsoft Academic seems to have the most comprehensive coverage, except in *Physics & Mathematics*, where it found fewer of the citations (57%) than the other sources.

*Table 3. Percentage of citations found by each data source, relative to the total number of citations found overall and by broad areas.*

|  | N | % of citations found (relative to N) | | | | | |
|---|---|---|---|---|---|---|---|
|  |  | Google Scholar | Microsoft Academic | Scopus | Dimensions | Web of Science | COCI |
| **Humanities, Literature & Arts** | 89,337 | 87% | 39% | 31% | 29% | 25% | 18% |
| **Social Sciences** | 406,661 | 88% | 47% | 40% | 36% | 33% | 20% |
| **Business, Economics & Management** | 235,338 | 88% | 47% | 34% | 32% | 29% | 19% |
| **Engineering & Computer Science** | 691,164 | 88% | 63% | 61% | 54% | 48% | 30% |
| **Physics & Mathematics** | 317,320 | 90% | 57% | 64% | 59% | 59% | 36% |
| **Health & Medical Sciences** | 1,001,507 | 85% | 63% | 59% | 58% | 51% | 27% |
| **Life Sciences & Earth Sciences** | 571,817 | 89% | 68% | 64% | 63% | 60% | 32% |
| **Chemical & Material Sciences** | 253,990 | 90% | 69% | 75% | 72% | 72% | 32% |

Further disaggregating the data to identify the percentage of relative citation overlap for each pair of sources in each subject area (Table 4), the patterns for the complete dataset (Figure 1) recur. Google Scholar consistently found most citations found by the other sources across all areas; there is a higher relative overlap between Microsoft Academic and Dimensions/COCI than between Microsoft Academic and Scopus/WoS; conversely, the relative overlap between Scopus and WoS is always higher than between Scopus and other sources; the highest relative overlap in each area is always for Dimensions/COCI; Microsoft Academic seems to lack coverage in *Physics & Mathematics*, as evidenced by its lower relative overlap in this area.

*Table 4. Relative pairwise overlaps between data sources (%). Overall and by broad subject areas.*

**A. Humanities, Literature & Arts**

| … that are also found by ⇒ Percentage of citations in ⇩ … | Google Scholar | Microsoft Academic | Scopus | Dimensions | Web of Science | COCI |
|---|---|---|---|---|---|---|
| Google Scholar | | 39% | 33% | 30% | 29% | 19% |
| Microsoft Acad. | 86% | | 57% | 62% | 53% | 42% |
| Scopus | 84% | 68% | | 65% | 68% | 42% |
| Dimensions | 89% | 86% | 75% | | 69% | 59% |
| Web of Science | 87% | 73% | 80% | 70% | | 46% |
| COCI | 93% | 92% | 77% | 94% | 73% | |

**B. Social Sciences**

| … that are also found by ⇒ Percentage of citations in ⇩ … | Google Scholar | Microsoft Academic | Scopus | Dimensions | Web of Science | COCI |
|---|---|---|---|---|---|---|
| Google Scholar | | 48% | 41% | 39% | 37% | 22% |
| Microsoft Acad. | 88% | | 66% | 69% | 60% | 40% |
| Scopus | 89% | 78% | | 75% | 76% | 43% |
| Dimensions | 93% | 90% | 83% | | 76% | 54% |
| Web of Science | 92% | 82% | 88% | 81% | | 47% |
| COCI | 96% | 95% | 85% | 96% | 80% | |

**C. Business, Economics & Management**

| … that are also found by ⇒ Percentage of citations in ⇩ … | Google Scholar | Microsoft Academic | Scopus | Dimensions | Web of Science | COCI |
|---|---|---|---|---|---|---|
| Google Scholar | | 46% | 35% | 34% | 31% | 20% |
| Microsoft Acad. | 85% | | 58% | 61% | 52% | 36% |
| Scopus | 91% | 80% | | 77% | 75% | 45% |
| Dimensions | 93% | 90% | 82% | | 75% | 55% |
| Web of Science | 93% | 84% | 87% | 83% | | 50% |
| COCI | 94% | 92% | 83% | 95% | 78% | |

**D. Engineering & Computer Science**

| … that are also found by ⇒ Percentage of citations in ⇩ … | Google Scholar | Microsoft Academic | Scopus | Dimensions | Web of Science | COCI |
|---|---|---|---|---|---|---|
| Google Scholar | | 65% | 62% | 58% | 55% | 32% |
| Microsoft Acad. | 90% | | 79% | 78% | 70% | 43% |
| Scopus | 89% | 82% | | 81% | 79% | 45% |
| Dimensions | 93% | 91% | 91% | | 82% | 53% |
| Web of Science | 93% | 86% | 94% | 87% | | 49% |
| COCI | 94% | 94% | 92% | 97% | 83% | |

*Table 4 (cont.) Relative pairwise overlaps between data sources. Overall and by broad subject areas.*

**E. Physics & Mathematics**

| … that are also found by ⇨<br>Percentage of citations in ⇩ … | Google Scholar | Microsoft Academic | Scopus | Dimensions | Web of Science | COCI |
|---|---|---|---|---|---|---|
| Google Scholar | | 58% | 65% | 61% | 61% | 37% |
| Microsoft Acad. | 91% | | 83% | 83% | 78% | 48% |
| Scopus | 91% | 74% | | 85% | 87% | 52% |
| Dimensions | 93% | 80% | 93% | | 88% | 60% |
| Web of Science | 93% | 75% | 95% | 88% | | 55% |
| COCI | 92% | 77% | 94% | 98% | 90% | |

**F. Health & Medical Sciences**

| … that are also found by ⇨<br>Percentage of citations in ⇩ … | Google Scholar | Microsoft Academic | Scopus | Dimensions | Web of Science | COCI |
|---|---|---|---|---|---|---|
| Google Scholar | | 64% | 61% | 62% | 58% | 29% |
| Microsoft Acad. | 87% | | 78% | 84% | 75% | 41% |
| Scopus | 88% | 84% | | 86% | 84% | 40% |
| Dimensions | 91% | 91% | 86% | | 82% | 45% |
| Web of Science | 95% | 87% | 92% | 89% | | 43% |
| COCI | 94% | 96% | 89% | 99% | 86% | |

**G. Life Sciences & Earth Sciences**

| … that are also found by ⇨<br>Percentage of citations in ⇩ … | Google Scholar | Microsoft Academic | Scopus | Dimensions | Web of Science | COCI |
|---|---|---|---|---|---|---|
| Google Scholar | | 69% | 67% | 67% | 64% | 34% |
| Microsoft Acad. | 91% | | 82% | 86% | 80% | 45% |
| Scopus | 93% | 88% | | 88% | 88% | 46% |
| Dimensions | 94% | 93% | 90% | | 87% | 50% |
| Web of Science | 95% | 91% | 94% | 91% | | 48% |
| COCI | 96% | 96% | 92% | 98% | 90% | |

**H. Chemical & Material Sciences**

| … that are also found by ⇨<br>Percentage of citations in ⇩ … | Google Scholar | Microsoft Academic | Scopus | Dimensions | Web of Science | COCI |
|---|---|---|---|---|---|---|
| Google Scholar | | 71% | 78% | 75% | 75% | 34% |
| Microsoft Acad. | 93% | | 90% | 92% | 88% | 43% |
| Scopus | 93% | 83% | | 89% | 92% | 40% |
| Dimensions | 94% | 89% | 94% | | 91% | 44% |
| Web of Science | 94% | 84% | 96% | 90% | | 41% |
| COCI | 95% | 93% | 93% | 98% | 91% | |

### 3.2.2. Full overlap

The differences in coverage between the older (Google Scholar, Scopus, WoS) and newer (Microsoft Academic, Dimensions) sources across subject areas are also evident from three-way comparisons (Figures 4, 6, and 8). The three-set combinations of data sources that are not displayed here are accessible from Appendix 1. The combinations that include more than one of the older sources are not included here because they were discussed in a previous study (Martín-Martín et al., 2018) and the results have barely changed. The combinations that include COCI are not displayed here because it is essentially a subset of the other sources (especially Dimensions).

Venn diagrams for the 252 specific subject categories are also accessible from Appendix 1. Figures 5, 7, 9 and 10 display the distribution of the proportions of citations that would fall in each section of the Venn diagrams calculated at this level of aggregation, for various pairs of data sources. The remaining combinations are accessible from Appendix 2.

*Google Scholar and the new sources: Microsoft Academic, and Dimensions*

For Google Scholar, Microsoft Academic, and Dimensions, the largest percentages of full overlap (citations found by the three sources) occur in the STEM fields (Figure 4). These range from 46% in *Physics and Mathematics*, to 63% in *Chemical and Material Sciences.* Full overlap in the areas of Humanities and Social Sciences is distinctly lower (25%-34%). This is caused by lower coverage of these areas in Microsoft Academic and Dimensions. The percentage of citations in Microsoft Academic and/or Dimensions that is not covered by Google Scholar ranges from 6% (in *Chemical and Material Sciences*) to 11% (in *Health & Medical Sciences*).

At the level of specific subject categories, for pairwise comparisons between Google Scholar/Microsoft Academic and Google Scholar/Dimensions (Figure 5) the general trend of the subject area is followed, with variations in some subject categories. The variation seems to be higher between Google Scholar/Dimensions than between Google Scholar/Microsoft Academic. Nevertheless, in both comparisons the percentages in the sector "Only in Google Scholar" are higher in the Humanities and Social Sciences, and lower in STEM fields. The sector "In both data sources" almost mirrors the one above, and the sectors "Only in Microsoft Academic" and "Only in Dimensions" have values almost exclusively below 10%, with two major exceptions. These correspond to the categories *Astronomy & Astrophysics*[17], and *Psychology*[18]. In these two categories, many citations found by Microsoft Academic and Dimensions were not found by Google Scholar. In the case of Psychology, the low citation coverage in Google Scholar is caused by one extremely highly-cited document (*Using thematic analysis in psychology*, by Virginia Braun and Victoria Clarke [19]), which at the time of data collection had 54,323 citations in Google Scholar. However, because of the limitations of Google Scholar's search interface for data extraction, only 10,996 citations could be extracted from Google Scholar for this article.

---

[17] Google Scholar/Microsoft Academic: https://osf.io/g8z42/; Google Scholar/Dimensions: https://osf.io/bwv5s/
[18] Google Scholar/Microsoft Academic: https://osf.io/jqwah/; Google Scholar/Dimensions: https://osf.io/xnf24/
[19] https://www.tandfonline.com/doi/abs/10.1191/1478088706QP063OA

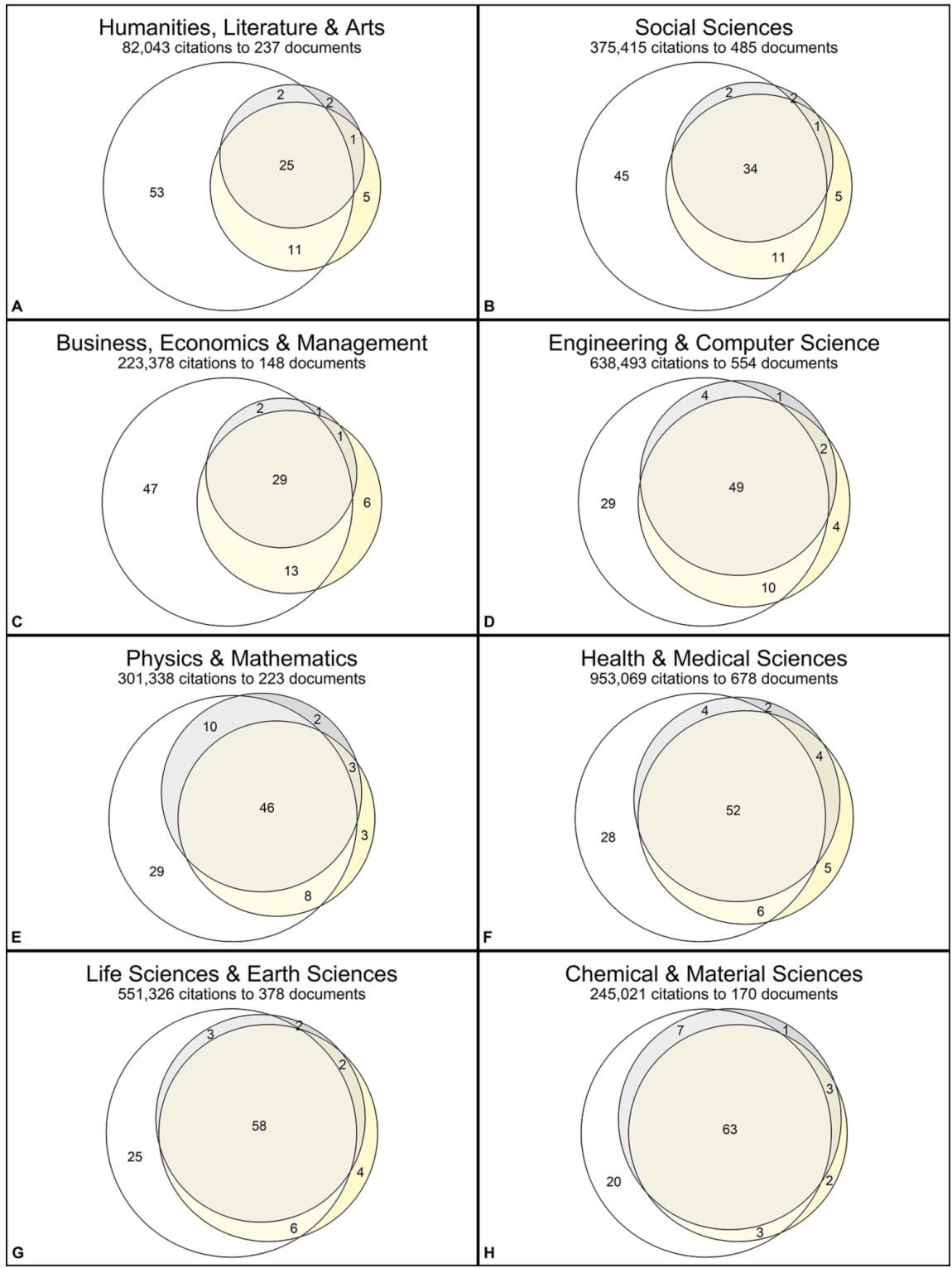

*Figure 4. Overlaps between citations found by Google Scholar, Microsoft Academic, and Dimensions in broad subject areas. Figures within Venn diagrams expressed as percentages.*

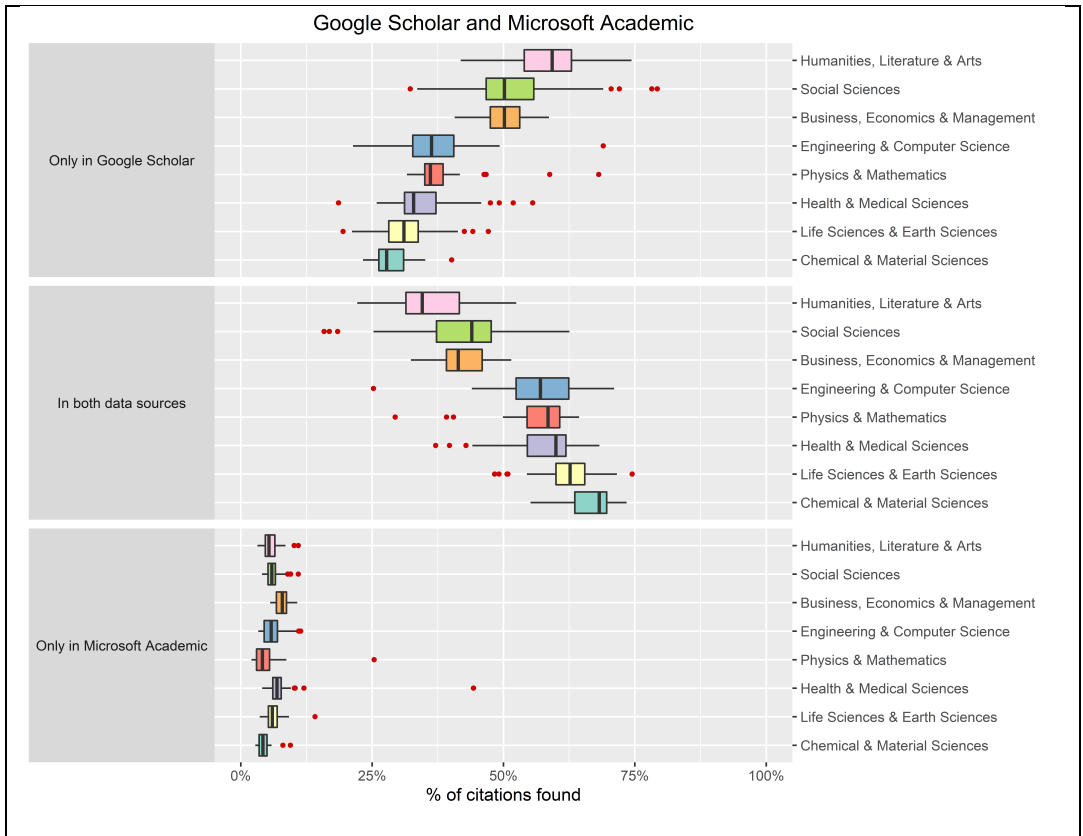

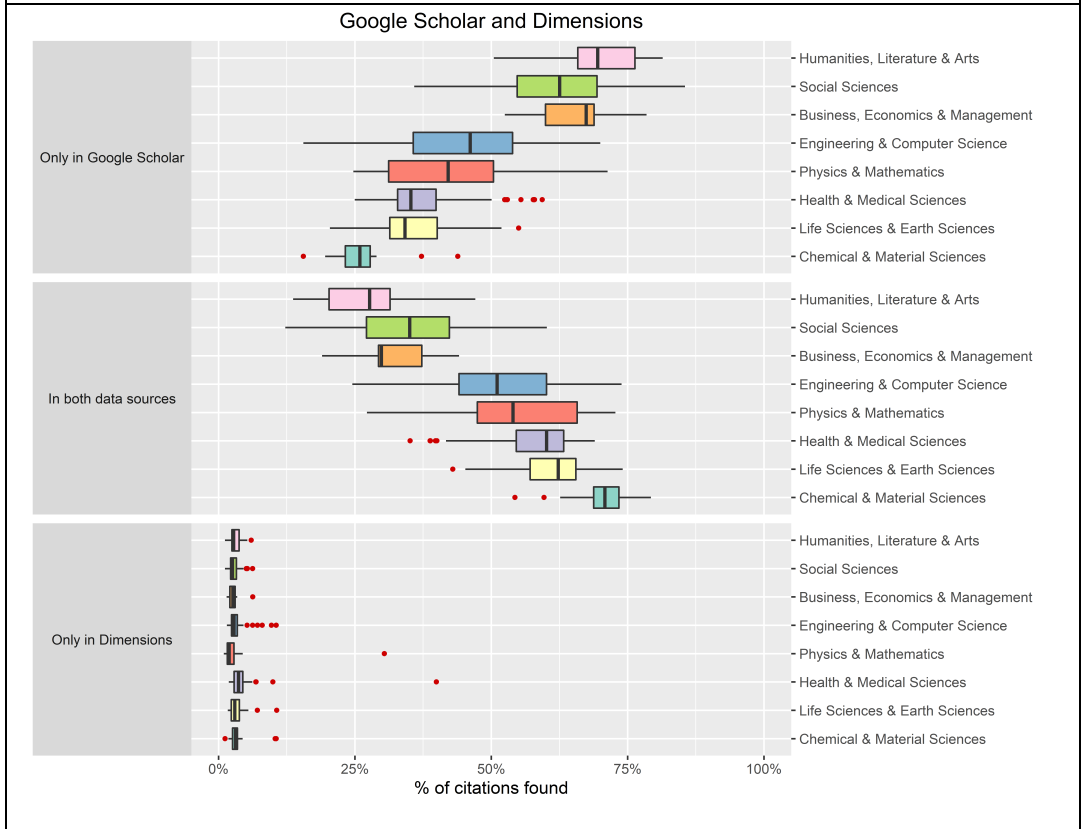

*Figure 5. Distribution of citations that fall within each sector of the Venn diagrams that compare Google Scholar to Microsoft Academic and Dimensions. Calculated at the level of subject categories, and aggregated by subject areas.*

*Scopus and the new sources: Microsoft Academic and Dimensions*

For Microsoft Academic, Scopus, and Dimensions, none of the sources is always larger than the others, with the results varying by subject area (Figure 6). Microsoft Academic sometimes has larger coverage than Scopus (Humanities and Social Sciences), although in these areas both contribute many unique citations. Scopus also sometimes provides more coverage than Microsoft Academic (*Physics & Mathematics*, *Chemical & Material Sciences*). The previously seen trend of higher percentages of full overlap in STEM fields also occurs here. The number of citations found by Dimensions is similar to that of Scopus across all subject areas, but there are also many citations that one of them finds that the other does not in the Humanities and Social Sciences. Comparing the three sources together, Dimensions provides the fewest unique citations.

In most subject categories (Figure 7), there are large Microsoft Academic/Scopus and Scopus/Dimensions citation overlaps. This is especially evident in STEM categories, where the overlap in almost all cases exceeds 50%. For Microsoft Academic/Scopus (Figure 7, top), there are 66 (out of 252) subject categories where the overlap exceeds 70%, and for Scopus/Dimensions, 148 categories exceed this overlap. Extreme cases of low overlap between sources are almost always in the Humanities and Social Sciences. For Microsoft Academic/Scopus, the lowest overlaps (below 30%) are in French Studies [20] (9%, although in this case the results are based only on citations to one seed document, because the rest were not covered by Microsoft Academic and Scopus), International Law [21] (20%), European Law [22] (21%), American Literature & Studies [23] (24%), Law [24] (26%), and Film [25] (27%). In 182 categories (out of 252) Microsoft Academic found more citations than Scopus. There are also some outlier cases of low overlap in STEM categories, such as over 50% of citations in Computer Graphics [26] and Discrete Mathematics [27] only being available in Microsoft Academic (compared to Scopus), or 48% of citations in High Energy & Nuclear Physics [28] and Quantum Mechanics [29] only being found by Scopus (compared to Microsoft Academic). For Scopus/Dimensions (Figure 7, bottom), many of the same categories have the lowest overlap: French Studies [30], International Law [31], American Literature & Studies [32], European Law [33], and History [34]. These low coverage figures are caused by Microsoft Academic and Dimensions having a lower coverage of citations in these categories than Scopus. In 36 categories (out of 252) Dimensions found more citations than Scopus.

---

[20] https://osf.io/gmrju/
[21] https://osf.io/bzha2/
[22] https://osf.io/f36sn/
[23] https://osf.io/7qzmk/
[24] https://osf.io/4gtdc/
[25] https://osf.io/ctzb7/
[26] https://osf.io/rz4cj/
[27] https://osf.io/v6bgy/
[28] https://osf.io/vafzp/
[29] https://osf.io/87cdh/
[30] https://osf.io/bqpz4/
[31] https://osf.io/p26ua/
[32] https://osf.io/fngph/
[33] https://osf.io/pdnxt/
[34] https://osf.io/xjhfw/

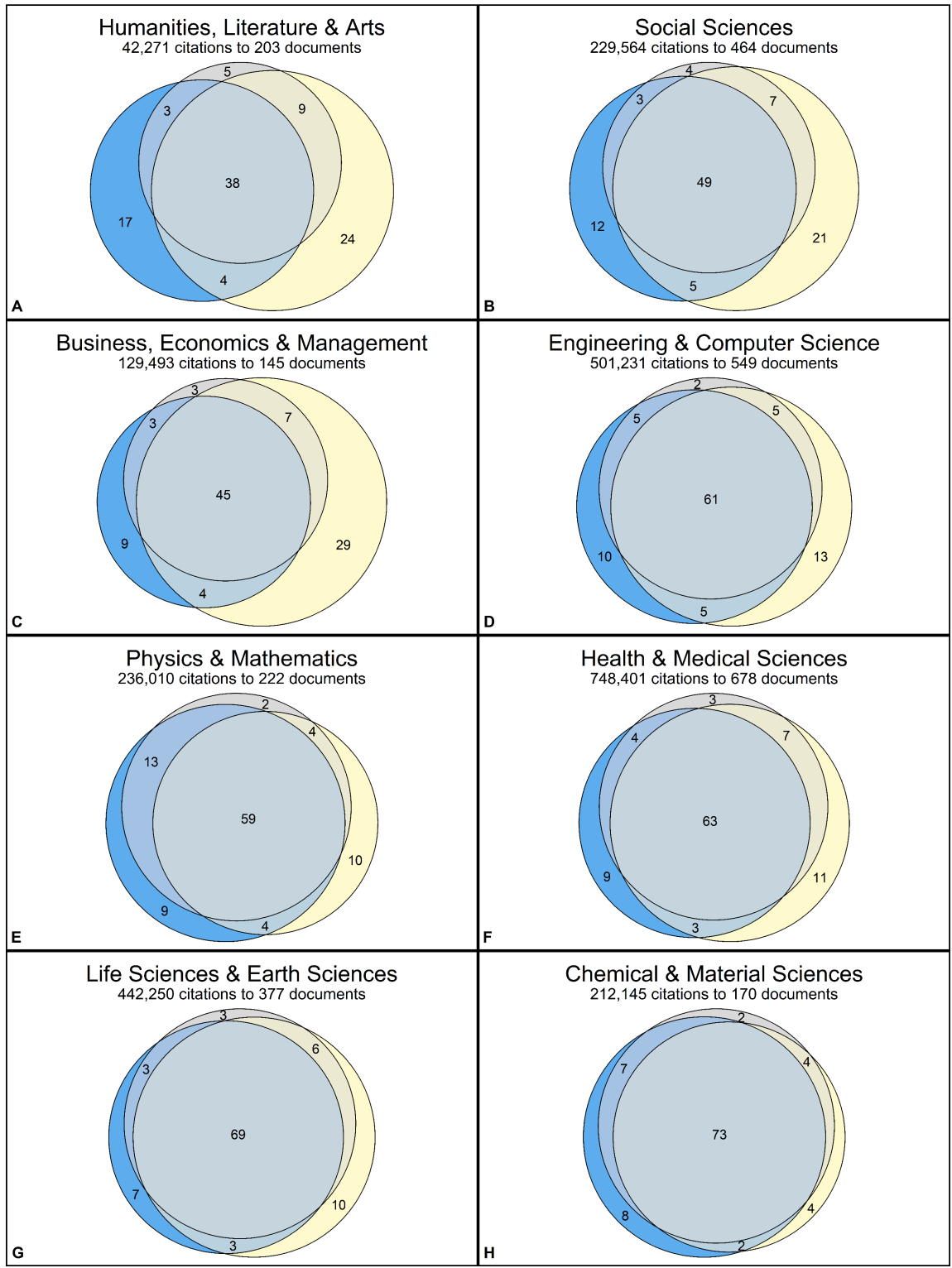

Figure 6. Overlap between citations found by Scopus, Microsoft Academic, and Dimensions, by broad subject area. Figures within Venn diagrams expressed as percentages.

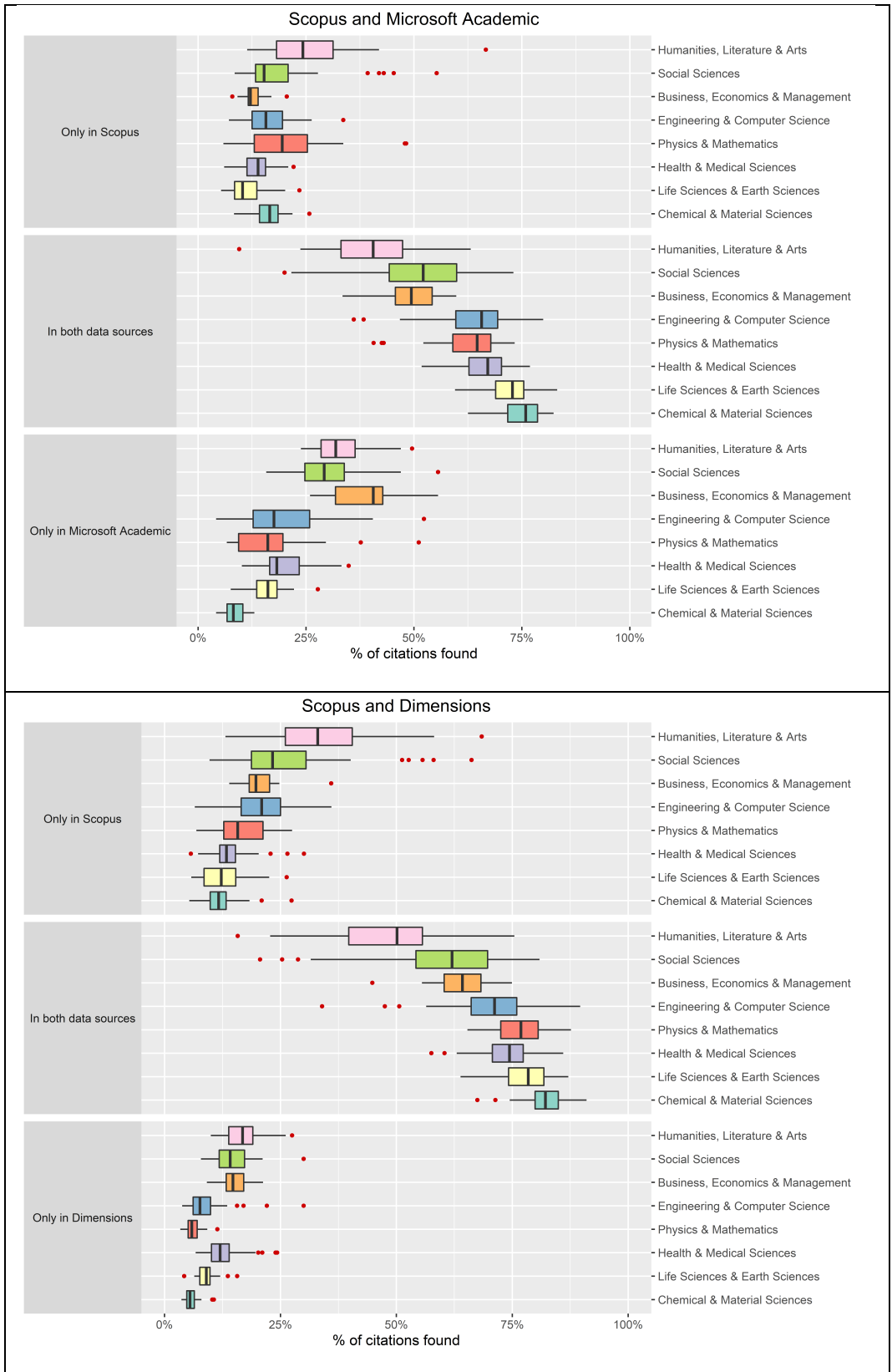

*Figure 7. Distribution of citations within each sector of the Venn diagrams that compare Scopus to Microsoft Academic and Dimensions. Calculated at the level of subject categories, and aggregated by subject areas.*

*Web of Science and the new sources: Microsoft Academic and Dimensions*

Comparing Microsoft Academic, Dimensions and WoS (Figure 8), there are many unique citations in Microsoft Academic and WoS. Out of these three, Dimensions found the fewest unique citations (2-6% depending on the area). Again, the divergence is higher in the Humanities and Social Sciences, where Microsoft Academic has the highest percentages of unique citations. Microsoft Academic also has lower coverage in *Physics & Mathematics* and (to a lower degree) in *Chemical & Material Sciences*.

The results by subject category confirm that some categories deviate from the trend in a broad area (Figure 9). Considering Microsoft Academic/WoS (Figure 9, top), Microsoft Academic's coverage is large compared to WoS for *Computing Systems*[35] (73% of all citations), *Software Systems*[36] (63%), *Educational Administration*[37] (62%), *Chinese Studies & History*[38] (60%), and *Discrete Mathematics*[39] (58%). The gaps in coverage of Microsoft Academic in *International Law*[40], and *Law*[41] occur again here, as 47% and 46% of the citations in these categories (respectively) are only found by WoS. Something similar occurs in the categories included in *Physics & Mathematics*: the distribution of citations only found by WoS in this area has an unusually wide interquartile range when compared with the other areas, which is a sign that Microsoft Academic's gaps in coverage in this area affect more than one category. The most extreme cases are again *Quantum Mechanics*[42] and *High Energy & Nuclear Physics*[43], with 47% and 44% of citations only found by WoS (respectively). In 223 categories (out of 252) Microsoft Academic found more citations than WoS. For the distributions of overlap and unique citations between Dimensions/WoS (Figure 9, bottom), there are some similarities with the previous comparison: 51% of the citations in *Computing Systems*[44] are only found by Dimensions, and in Humanities and Social Sciences over a third of the citations in *Chinese Studies & History*[45], and *Foreign Language Learning*[46] are only found by Dimensions, which reveals coverage gaps in these categories in WoS. In other Humanities categories, such as *American Literature & Studies*[47] (51%), *History*[48] (46%), or *Literature & Writing*[49] (46%) WoS found more unique citations than Dimensions. Dimensions also has gaps in coverage in *Computer Graphics*[50], *International Law*[51], *Law*[52], and *Middle Eastern & Islamic Studies*[53], compared to WoS. In 185 categories (out of 252) Dimensions found more citations than WoS.

---

[35] https://osf.io/ugvh3/
[36] https://osf.io/6vrnp/
[37] https://osf.io/x9g3e/
[38] https://osf.io/54xky/
[39] https://osf.io/fa8sr/
[40] https://osf.io/9584j/
[41] https://osf.io/h7jt2/
[42] https://osf.io/ghws2/
[43] https://osf.io/gpyse/
[44] https://osf.io/rsj4m/
[45] https://osf.io/bvr3p/
[46] https://osf.io/vmdbx/
[47] https://osf.io/zd53e/
[48] https://osf.io/q529p/
[49] https://osf.io/qcdsh/
[50] https://osf.io/sfd2g/
[51] https://osf.io/a9mtx/
[52] https://osf.io/n2e98/
[53] https://osf.io/za5ks/

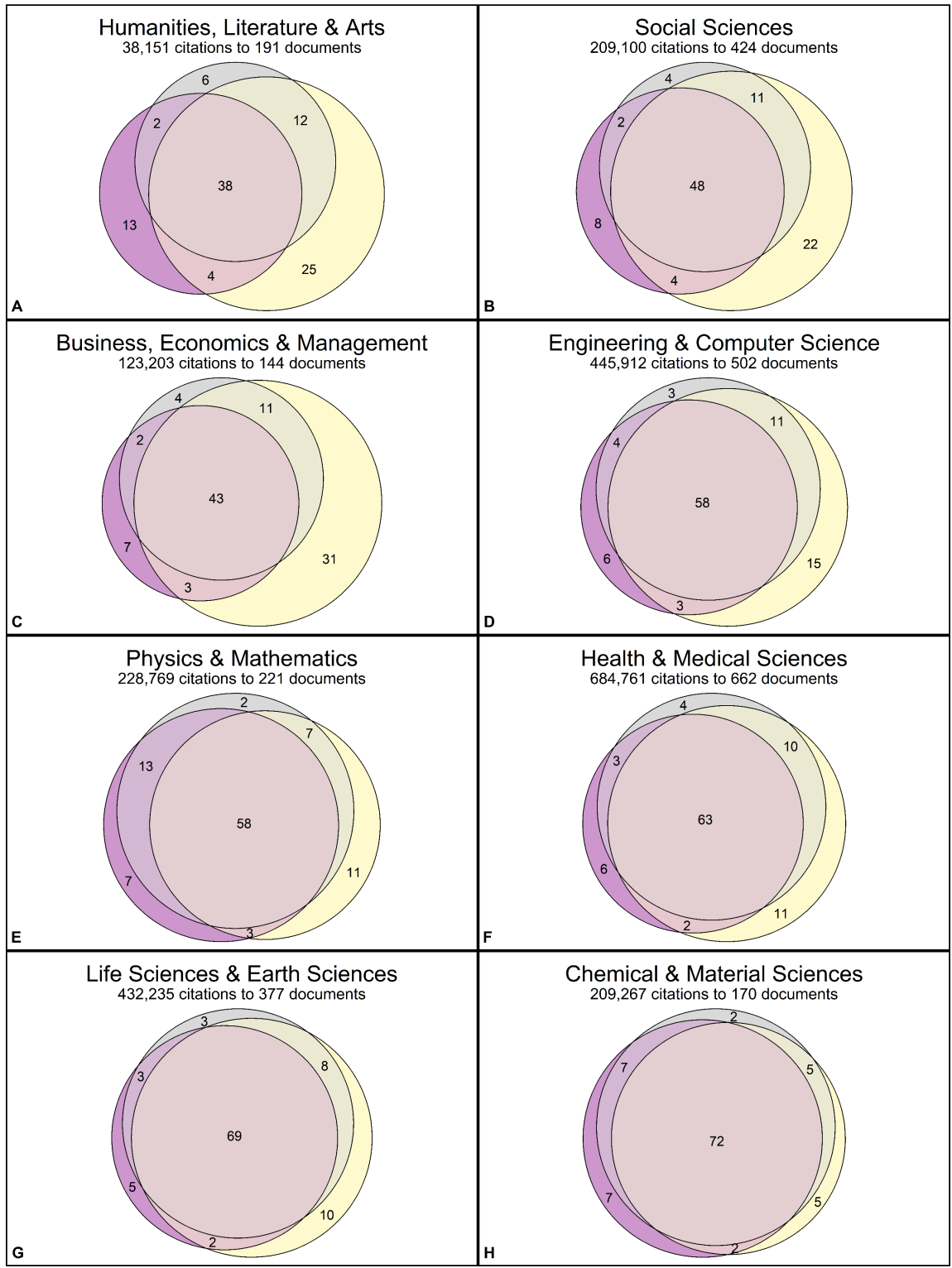

*Figure 8. Overlaps between citations found by Web of Science, Microsoft Academic, and Dimensions, by broad subject areas. Figures within Venn diagrams expressed as percentages.*

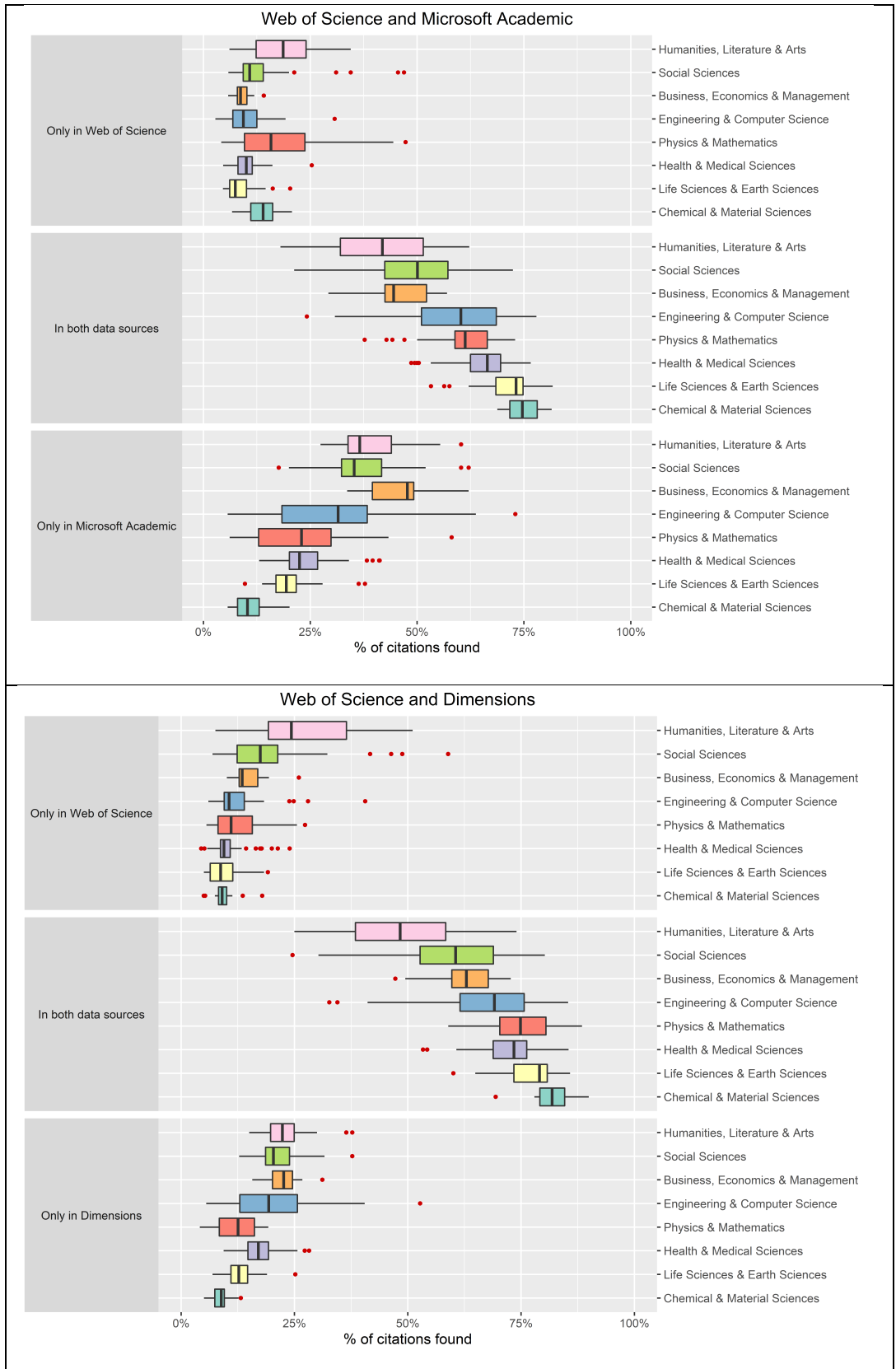

*Figure 9. Distribution of citations within each sector of the Venn diagrams that compare Web of Science to Microsoft Academic and Dimensions. Calculated at the level of subject categories, and aggregated by subject areas.*

*Microsoft Academic and Dimensions*

At the level of subject categories, the vast majority of citations in Microsoft Academic/Dimensions are found either by both databases, or only by Microsoft Academic. In 209 out of 252 categories, the percentage of unique citations in Dimensions is below 10% (Figure 10). The exceptions are in *Physics & Mathematics*, where 45% of the citations in *Quantum Mechanics*[54], 39% of the citations in *High Energy & Nuclear Physics*[55], and 26% of the citations in *Plasma & Fusion*[56] (also included in *Engineering & Computer Science*) are only found by Dimensions. This again reveals the gap in coverage of Microsoft Academic in these categories. In 226 categories (out of 252) Microsoft Academic found more citations than Dimensions.

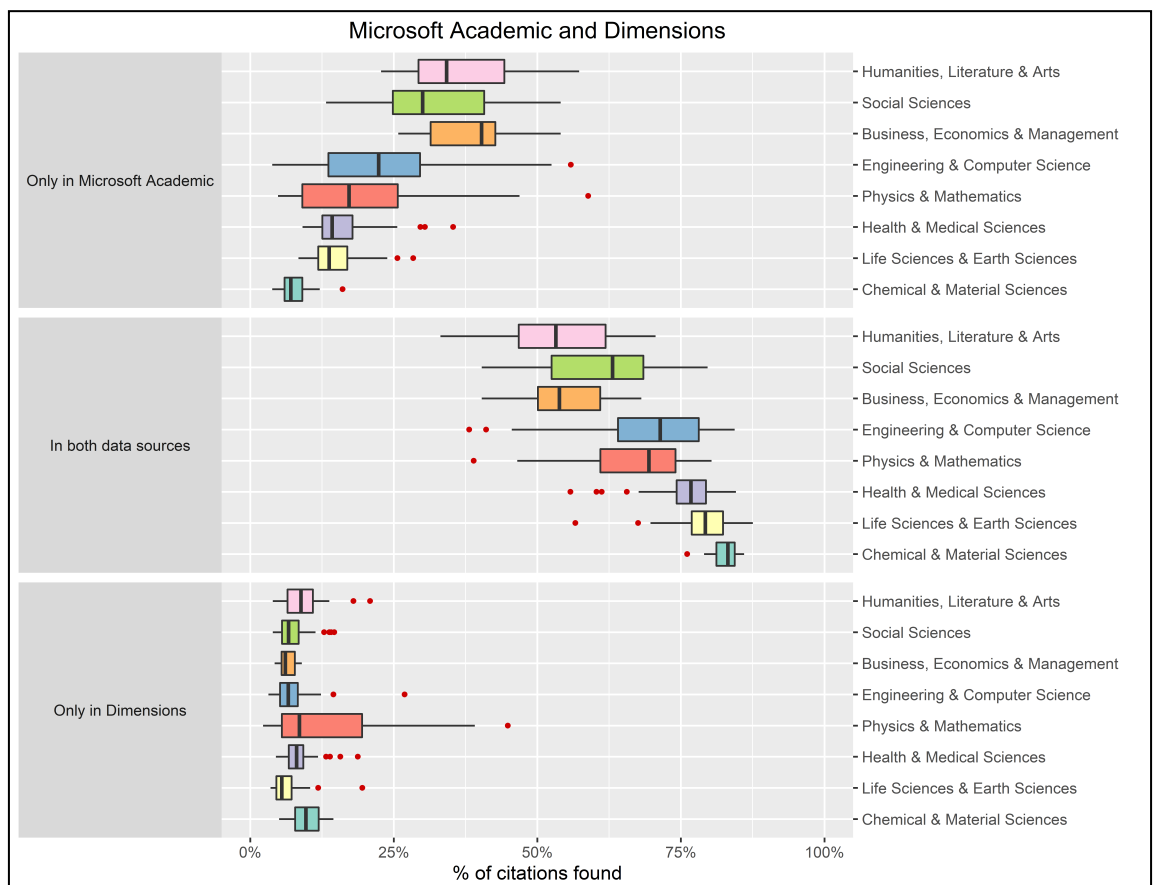

*Figure 10. Distribution of citations within each sector of the Venn diagrams that compare Microsoft Academic and Dimensions. Calculated at the level of subject categories, and aggregated by subject areas.*

# 4. Discussion

## 4.1. Limitations

Because this study uses an updated and extended version of the sample used in Martín-Martín et al. (2018), many of the limitations declared in that study are also applicable here, as summarised below.
- The seed sample of documents used all highly-cited documents published in English in 2006. To generalize the results presented here, it must be assumed that the population of documents that cite these highly-cited documents is comparable to the general population of citing documents within each subject category. This might not be true in some cases, because different topics within the same category might have different citation patterns (certain highly-cited topics within a category might be overrepresented). Also, these results probably do not represent the reality of coverage of academic literature published in languages other than English and literature about regionally-relevant topics, where Google Scholar, Microsoft Academic, Dimensions and COCI may have an advantage.

---
[54] https://osf.io/3npwu/
[55] https://osf.io/7qb8v/
[56] https://osf.io/n5j8v/

- Google Scholar might have an unfair advantage in this analysis, as the initial seed sample was selected from a list of the highest-cited documents in this source (the accuracy of citation detection of Google Scholar in this specific sample could be higher than the average accuracy of citation detection across all documents in Google Scholar, which is unknown). However, the correlations between the citation counts of citing documents available in Martín-Martín et al. (2018) suggest that this advantage is not substantial: when analysing documents from the entire distribution of citation counts (not only highly-cited documents), Google Scholar still consistently reports higher citation counts than WoS and Scopus, while providing essentially the same citation rankings at the document level in most subject categories as the other two sources.
- The algorithm that matches citations across data sources is intentionally conservative: it is set to minimize false positives, potentially at the expense of false negatives. Therefore, the percentages of overlap shown in this study are lower bounds.
- Unlike Martín-Martín et al. (2018), where citations from documents included in the ESCI Backfile for documents published between 2005 and 2014 were not included in the analysis, in this study all available citation data available in the citation indexes that are part of WoS Core Collection is analysed.
- Data extraction for this analysis was carried out in May/June 2019. However, the rapid development of these platforms may render the results obsolete in the future. Updated analyses similar to this one might be necessary to ascertain the current coverage of the data sources, especially if regular reports on coverage development are not issued by the sources themselves.

Other aspects related to coverage, such as the distribution of document types, language, date of publication, or indexing speed are not analysed here and could be investigated in future studies, as they are also necessary for users who need to decide which data source(s) are most suitable for their needs.

### 4.2. Comparison with previous studies

The results generally agree with previous studies comparing the coverage of Microsoft Academic and Dimensions. Similarly to Harzing & Alakangas (2017a, 2017b) and Thelwall (2017), Microsoft Academic detected more citations than WoS and Scopus. This citation detection advantage seems to be higher in the Humanities, Social Sciences, and Business & Economics than in the other areas, where in some cases Microsoft Academic had lower coverage (Physics, Chemistry). The results here cannot be directly compared to Hug & Brändle (2017), who reported that Scopus had slightly greater coverage of journal articles than Microsoft Academic, because this study does not analyse specific document types. However, assuming that most citations come from journal articles, Microsoft Academic seems to have now surpassed Scopus in raw size, at least in the three areas mentioned above.

For Dimensions, the results also agree with those reported by Harzing (2019), who found that it had a similar or better coverage than WoS and Scopus in Business & Economics. Here the results show that the three data sources offer a similar coverage (Scopus is slightly larger, followed by Dimensions), but each can detect a non-negligible proportion of citations that the others can't.

From Visser et al. (2020) the percentage of documents covered by Scopus that are also covered by Dimensions is 78%, but in this study the percentage of citations found by Scopus that are also found by Dimensions is higher (84%). The causes of the difference between these figures is unclear, but some possibilities are a) this study uses a sample of citations while Visser et al. use the entire collection of source documents, b) the possibility that Dimensions has a lower coverage of older documents (our study analyses citations from 2006-2018, while Visser et al. analysed coverage between 1996-2017), or c) that there was an increase in coverage between the time Visser et al. obtained their data (December 2018), and the time the data for this study was extracted (May-June 2019). The overlap Visser et al. found between Scopus and WoS is significantly lower than found here: according to their results (overlap of 29.1 million documents, and 44.9 million documents in total in Scopus), WoS covered 65% of the documents available in Scopus. In the current study, however, WoS found 83% of the citations found by Scopus. The cause of this significant difference is also unknown, but it might be in part caused by the fact that Visser et al. analysed only documents in the SCI, SSCI, and A&HCI and the Conference Proceedings Citation Index (CPCI), while this study also considers other citation indexes within WoS Core Collection, such as ESCI and BKCI. Lastly, the percentage of Scopus documents also covered by Microsoft Academic reported by Visser et al. (81%) is very similar to the percentage of Scopus citations also found by Microsoft Academic reported here (82%). However, the full overlap between the two sources is much higher here (66%) than in Visser et al. (18%), because in the latter study a much higher amount of unique content was detected in Microsoft Academic. One possible reason for this might be that our study only considers documents with recorded relationships to other documents (through citations), while some of the documents in Microsoft Academic analysed in Visser et al. might not have these connections, which would make them undetectable to our methodology.

Although most of the results of the overlap analysis reported here closely match those of the previous study with the same seed set (Martín-Martín et al., 2018), several discrepancies were found. In two subject categories (Psychology, and Astronomy & Astrophysics), the updated analysis showed that Google Scholar had a lower coverage than the other data sources, while in the old dataset, this was not the case. In the case of Astronomy & Astrophysics, this apparent fluctuation in coverage is consistent with an editorial published in August of 2019 in the journal Astronomy & Astrophysics, which denounced a sharp decrease in the h5-index of this journal in the 2019 edition of Google Scholar Metrics Forveille (2019) caused by a technical error in Google Scholar. Therefore, this seems to be a new case of a major coverage outage in Google Scholar, similar to one previously reported by Delgado López-Cózar & Martín-Martín (2018) which affected many journals published in Spain, and which was resolved when Google Scholar rebuilt its index a few months later. This issue will be analysed in detail in a future study as an example of how coverage in Google Scholar can suffer large (downward) fluctuations over time, as this can negatively affect literature search.

# 5. Conclusions

## 5.1. Comprehensiveness of data sources across subject categories

The results show that Google Scholar is still the most comprehensive data source among the six studied here. This holds true for the overall results and the results across all subject areas, with some exceptions such as *Astronomy & Astrophysics*. Google Scholar found nearly all the citations found by Microsoft Academic, Dimensions, and COCI (89%, 93%, and 94%, respectively). The largest divergences occur in the Humanities and Social Sciences (lowest value is 84%, which corresponds to the percentage of Scopus citations in the Humanities found by Google Scholar). Additionally, there is a significant amount of extra coverage in Google Scholar that is not found in any of the other data sources (26% of all citations across all data sources). Google Scholar could therefore make an important contribution to the scientific community by opening its bibliographic and citation data, which would also facilitate the identification of errors such as coverage fluctuations.

Whilst the results confirm that Microsoft Academic and Dimensions provide at least as many citations as Scopus and WoS in many subject categories, some gaps still exist:

- Microsoft Academic seems to index the Humanities, Social Sciences, and Business, Economics & Management particularly well, although not for all categories.
- Dimensions is closely behind Scopus in all areas in terms of citations found, but surpasses WoS in all areas, except in two (Physics & Mathematics, and Chemical & Material Sciences) where they are tied, although there are also differences at the level of subject categories (Dimensions also has coverage gaps in some Humanities categories).

## 5.2. Implications for academic literature search

Although Google Scholar and Microsoft Academic are the two most comprehensive bibliographic data sources analysed in this study, their search functionalities have a number of limitations, such as limited support of Boolean and other types of search operators, limited filtering capabilities (Google Scholar), and non-transparent algorithms to process queries and rank the documents in the results page (Microsoft Academic uses artificial intelligence, and Google Scholar uses publicly unknown heuristics to rank documents by relevance) (Beel & Gipp, 2009c, 2009b, 2009a; Martin-Martin et al., 2017; Orduña-Malea et al., 2016; Rovira et al., 2019; Wang et al., 2020).These characteristics, which prevent users from being able to generate complex search equations that are guaranteed to stay reproducible over time, have led some authors to consider Google Scholar and Microsoft Academic inadequate for query-based search (Gusenbauer & Haddaway, 2020). Dimensions, which does not allow complex Boolean searches in its web interface either, was not analysed in that study.

On the other hand, Scopus and WoS have a lower coverage, especially in some areas such as the Humanities and Social Sciences, do not cover non-peer-reviewed scientific documents (Martín-Martín et al., 2018), are slower at indexing (Moed et al., 2016), and are not free. These characteristics reduce their usefulness in situations where fast and unrestricted access to the latest studies is important, such as the COVID-19 pandemic in which preprints play a critical role (Fraser et al., 2020). Nevertheless, these sources offer advanced search and filtering functionalities, and were considered suitable tools for evidence synthesis in the form of systematic reviews (Gusenbauer & Haddaway, 2020).

Thus, there seems to be a mismatch between the bibliographic data sources that are currently the most comprehensive, and those that offer users the most control over their searches. Since systematic reviews require both comprehensiveness of coverage and control over the search process, it is possible that in some cases no single currently available data source is adequate for the task, and instead at least two sources should be used. One way to do this would be to expand the concept of systematic search beyond the traditional search query to include other non-query-based search processes that can also be carried out in a systematic and reproducible manner. One possibility would be the expansion of a document collection obtained in a query-based search through the analysis of its citation network. This expansion can be carried out in a more comprehensive data source, different from the one where the initial search was carried out. As a longer term solution, academic search tools should strive to offer more a transparent and reproducible search process and embrace community standards for interoperability and reuse of document metadata (Haddaway & Gusenbauer, 2020).

Lastly, searches suitable for systematic reviews are only one of the many types of search that are carried out in these data sources. Indeed, the more recent academic search platforms (Microsoft Academic, Semantic Scholar, Dimensions) have not implemented traditional advanced query-based capabilities (Dimensions supports them in its API), and seem to be instead focusing on the browsing experience (advanced filtering), and in offering analytics dashboards. Lens.org seems to be an exception, as it also offers advanced structured query-based search (Tay, 2019). Future studies could focus on the suitability of these and other bibliographic data sources to solve specific types of information needs, as it is important that researchers are aware of the strengths and limitations of each data source for specific use cases and in specific knowledge domains.

### 5.3. Implications for bibliometric analyses

As new sources of bibliographic data (including citation data) become openly available and validated for specific types of bibliometric analyses, the need to rely on expensive proprietary data sources may diminish. Regarding the findings in this study, the final decision about which source to use may depend on properties of the sources other than coverage, such as metadata quality and bulk access options. If these factors are not of overriding importance, however, then Google Scholar is the best choice in almost all subject areas for those needing the most comprehensive citation counts but not needing complete lists of citing sources. If complete lists are needed, then Microsoft Academic is the best alternative and is also free. The amount of citation data in the public domain (through COCI) is still low and not useful on its own, presumably because its role is to feed other sources, not to be more comprehensive than them.

In use cases where exhaustiveness of coverage is required, but coverage divergence is considered to be large (many unique citations in each data source), the combination of several sources is recommended.

To conclude, the evidence presented in this study is intended to serve as a tool for researchers and other users of bibliographic databases, one that will hopefully help them make more informed decisions when they need to select one or more of these data sources to solve a specific information need.

# 6. Acknowledgements

We thank Medialab UGR (Universidad de Granada) for providing funding to cover the cost of hosting the interactive web application [57] created to explore the data used in this study. We thank Digital Science for providing free access to the Dimensions API. We thank JingXuan Xie for translating the abstract to Chinese. We thank Asura Enkhbayar for suggesting the use of an upset plot in Figure 2. Lastly, we thank two anonymous reviewers for their thoughtful comments, which have helped improved the manuscript substantially.

---

[57] https://albertomartin.shinyapps.io/citation_overlap_2019/

# Appendix 1
Complete list of Venn diagrams computed for this study

**No subject aggregation**

| | |
|---|---|
| Two-set Venn diagrams (all subject categories) | https://osf.io/bwpaq/ |
| Three-set Venn diagrams (all subject categories) | https://osf.io/jkrge/ |

**Aggregated by 8 subject areas**

| | |
|---|---|
| Google Scholar – Microsoft Academic – Scopus | https://osf.io/h7m8s/ |
| Google Scholar – Microsoft Academic – Dimensions | https://osf.io/7v4kr/ |
| Google Scholar – Microsoft Academic – Web of Science | https://osf.io/fn3yh/ |
| Google Scholar – Microsoft Academic – COCI | https://osf.io/s3bmp/ |
| Google Scholar – Scopus – Dimensions | https://osf.io/q8ecx/ |
| Google Scholar – Scopus – Web of Science | https://osf.io/qkc2a/ |
| Google Scholar – Scopus – COCI | https://osf.io/mrvdb/ |
| Google Scholar – Dimensions – Web of Science | https://osf.io/nwm83/ |
| Google Scholar – Dimensions – COCI | https://osf.io/dzs5x/ |
| Google Scholar – Web of Science – COCI | https://osf.io/64chg/ |
| Microsoft Academic – Scopus – Dimensions | https://osf.io/hgzn6/ |
| Microsoft Academic – Scopus – Web of Science | https://osf.io/f7xpa/ |
| Microsoft Academic – Scopus – COCI | https://osf.io/c6tpz/ |
| Microsoft Academic – Dimensions – Web of Science | https://osf.io/f5zxs/ |
| Microsoft Academic – Dimensions – COCI | https://osf.io/ry87a/ |
| Microsoft Academic – Web of Science – COCI | https://osf.io/vxyj4/ |
| Scopus – Dimensions – Web of Science | https://osf.io/xqg3y/ |
| Scopus – Dimensions – COCI | https://osf.io/jmvb6/ |
| Scopus – Web of Science – COCI | https://osf.io/e43kt/ |
| Dimensions – Web of Science - COCI | https://osf.io/ew7fj/ |

**Aggregated by 252 subject categories (zipped)**

| | |
|---|---|
| Google Scholar – Microsoft Academic | https://osf.io/v4ek3/ |
| Google Scholar – Scopus | https://osf.io/umsyw/ |
| Google Scholar – Dimensions | https://osf.io/jqmuy/ |
| Google Scholar – Web of Science | https://osf.io/4b8uq/ |
| Google Scholar – COCI | https://osf.io/gytuh/ |
| Microsoft Academic – Scopus | https://osf.io/jw2bt/ |
| Microsoft Academic – Dimensions | https://osf.io/a2mp7/ |
| Microsoft Academic – Web of Science | https://osf.io/2hkxq/ |
| Microsoft Academic – COCI | https://osf.io/ch4gb/ |
| Scopus – Dimensions | https://osf.io/q4swk/ |
| Scopus – Web of Science | https://osf.io/qcpbh/ |
| Scopus – COCI | https://osf.io/2xvbh/ |
| Dimensions – Web of Science | https://osf.io/pdycb/ |
| Dimensions – COCI | https://osf.io/j7qte/ |
| Web of Science - COCI | https://osf.io/mnwe7/ |

# Appendix 2
Complete list of boxplots computed for this study

**Subject category-level overlap data aggregated by 8 subject areas**

| | |
|---|---|
| Google Scholar – Microsoft Academic | https://osf.io/b94xp/ |
| Google Scholar – Scopus | https://osf.io/rvbw9/ |
| Google Scholar – Dimensions | https://osf.io/ubtrm/ |
| Google Scholar – Web of Science | https://osf.io/7wb49/ |
| Google Scholar – COCI | https://osf.io/7ekdr/ |
| Microsoft Academic – Scopus | https://osf.io/jx7by/ |
| Microsoft Academic – Dimensions | https://osf.io/x4257/ |
| Microsoft Academic – Web of Science | https://osf.io/rdw7g/ |
| Microsoft Academic – COCI | https://osf.io/f8a9e/ |
| Scopus – Dimensions | https://osf.io/3a97k/ |
| Scopus – Web of Science | https://osf.io/w4zv3/ |
| Scopus – COCI | https://osf.io/jtnyu/ |
| Dimensions – Web of Science | https://osf.io/gsjwm/ |
| Dimensions – COCI | https://osf.io/sr4wu/ |
| Web of Science - COCI | https://osf.io/6dkw4/ |

# Resumen 

**Introducción**

Recientemente han aparecido nuevas fuentes de datos de citas, como Microsoft Academic, Dimensions, y el índice citas DOI-a-DOI con datos de CrossRef realizado por OpenCitations (COCI). Aunque estas fuentes ya han sido comparadas con Web of Science, Scopus, y Google Scholar, todavía no hay evidencias sistemáticas sobre sus diferencias a nivel de categorías temáticas.

**Metodología**

En respuesta, este trabajo analiza 3.073.353 citas encontradas por estas seis fuentes a 2.515 documentos altamente citados publicados en inglés en 2006, clasificados en 252 categorías temáticas, expandiendo y actualizando así el estudio con una mayor muestra publicado anteriormente.

**Resultados**

Google Scholar encontró el 88% de todas las citas, (muchas de las cuales no fueron detectadas por las otras fuentes) así como la mayoría de las citas encontradas por las otras fuentes (89%-94%). Este patrón se mantenía en la mayoría de las categorías temáticas. Microsoft Academic es la segunda fuente más grande (60% de todas las citas), incluyendo el 82% de las citas de Scopus y el 86% de las de Web of Science. En la mayoría de las categorías, Microsoft Academic encontró más citas que Scopus y Web of Science (en 182 y 223 categorías, respectivamente), pero tenía huecos en la cobertura de algunas áreas, como en Física y algunas categorías de las Humanidades. Después de Scopus, Dimensions es la cuarta fuente más grande (54% de todas las citas) incluyendo el 84% de las citas de Scopus y el 88% de las de Web of Science. Dimensions encontró más citas que Scopus en 36 categorías, más que Web of Science en 185, y también presenta algunos huecos de cobertura, especialmente en las Humanidades. Después de Web of Science, COCI es la fuente más pequeña, con el 28% de todas las citas.

**Conclusiones**

Google Scholar es todavía la fuente con mayor cobertura. En muchas categorías temática MA y Dimensions son ya buenas alternativas a Scopus y Web of Science en términos de cobertura.

# 摘要（回到首页）

**引言**

Microsoft Academic，Dimensions 和带有 OpenCitations (COCI)发布的 CrossRef 数据的 DOI-a-DOI 引文索引，作为新出现的引文数据库，尽管已经与 Web of Science，Scopus 和 Google Scholar 进行了比较，但是仍然没有研究验证它们在主题类别方面的差异，本文将对此进行系统性的研究。

**研究方法**

本文分析了这六种数据库的 3,073,353 个引用，引用来自 2006 年发表的 2515 篇英文高被引文章，文章归属 252 个主题类别。采用近期发表的更大样本数让研究更具概括性和及时性。

**结果**

Google Scholar 可以发现所有引用中的 88%（其中许多未被其他数据库检测到）以及其他来源中的大多数被引（89%-94%）。在大多数主题类别中都是如此。Microsoft Academic 是第二大数据来源（占所有引用的 60%），包括 82%的 Scopus 引用和 86%的 Web of Science 引用。在大多数类别中，Microsoft Academic 所引用的文献多于 Scopus 和 Web of Science（分别在 182 个和 223 个类别中），但在某些领域如物理和某些人文学科的覆盖范围上则表现较弱。Dimensions 是仅次于 Scopus 的第四大来源（占所有引用的 54%），包括 84%的 Scopus 引用和 88%的 Web of Science 引用，并在 36 个类别中的引用多于 Scopus，在 185 个类别中多于 Web of Science。Dimensions 在覆盖面上也有薄弱，特别是人文学科。COCI 是 Web of Science 之后，覆盖面最少的数据来源，占所有引用的 28%。

**结论**

Google Scholar 仍然是覆盖面最高的数据来源。在很多主题类别上 Microsoft Academic 和 Dimensions 是替代 Scopus 和 Web of Science 不错的选择。

## Publication history [(back to first page)](#)

| Version nº | Date | Changes |
|---|---|---|
| 1 | 29/04/2020 | Preprint [submitted to arXiv](#) |
|   | 30/04/2020 | Preprint submitted to *Scientometrics* for evaluation |
|   | 05/06/2020 | Editor decision: Major revision |
| 2 | 06/08/2020 | Revision of manuscript submitted to *Scientometrics* <ul><li>Various clarifications added to Introduction and Methodology</li><li>Figure 2 changed from a Euler diagram to an upset plot.</li><li>Minor cosmetic changes to figures and tables.</li><li>Discussion and Conclusions sections has been expanded.</li></ul> |
|   | 18/08/2020 | Editor decision: Accept |
|   | 21/09/2020 | Publisher-branded version [available from Springer](#) |
|   | 21/09/2020 | Authors' Accepted Manuscript is [submitted to arXiv](#), updating previously submitted preprint |
| 3 | 30/01/2021 | Article has been included in a completed journal issue (*Scientometrics,* 126(1), 871-906). PDF and metadata updated in [arXiv](#). |